\documentclass[conference]{IEEEtran}
\IEEEoverridecommandlockouts

\usepackage{graphicx}
\usepackage{epigraph}
\usepackage{amsmath,amssymb,amsfonts}
\usepackage{tikz}
\usetikzlibrary{arrows}
\usepackage{enumerate}
\usepackage{bm}
\usepackage{dsfont}
\usepackage{csquotes}
\usepackage{cite}
\usepackage{textcomp}
\usepackage{mathtools}
\usepackage{cleveref}
\usepackage{verbatim}
\usepackage{placeins}
\usepackage{listings}
\usepackage{caption}
\usepackage[nocomma]{optidef}
\usepackage{empheq}
\usepackage{algorithm,algorithmic}
\usepackage{subcaption}
\usepackage{xcolor}
\usepackage{bbm}

\usepackage{booktabs}
\usepackage{tabularx}
\usepackage{enumitem}

\usepackage{tabularx}

\usepackage{pgfplots}

\newcommand*\link[1]{\hspace*{0em plus 1fill}\makebox{(#1)}}

\def\BibTeX{{\rm B\kern-.05em{\sc i\kern-.025em b}\kern-.08em
    T\kern-.1667em\lower.7ex\hbox{E}\kern-.125emX}}

\begin{document}

\title{On Symbol Error Probability-based Beamforming in MIMO Gaussian Wiretap Channels
\thanks{}
}

\author{\IEEEauthorblockN{$\textnormal{Nam Nguyen}^{1, *}$, $\textnormal{An Vuong}^{1}$, $\textnormal{Thuan Nguyen}^{2}$, and $\textnormal{Thinh Nguyen}^{1}$}
\IEEEauthorblockA{$^{1}\textnormal{School of Electrical and Computer Engineering, Oregon State University, Corvallis, OR, 97331}$\\
$^{2}\textnormal{Department of Engineering, Engineering Technology, and Surveying, East Tennessee State University, Johnson, TN, 37604}$\\}
}

\maketitle 
\begin{abstract}
This paper investigates beamforming schemes designed to minimize the symbol error probability (SEP) for an authorized user while guaranteeing that the likelihood of an eavesdropper correctly recovering symbols remains below a predefined threshold. Unlike previous works that focus on maximizing secrecy capacity,  our work is centered around finding an optimal beamforming vector for binary antipodal signal detection in multiple-input multiple-output (MIMO) Gaussian wiretap channels. Finding the optimal beamforming vector in this setting is challenging. Computationally efficient algorithms such as convex techniques cannot be applied to find the optimal solution. To that end, our proposed algorithm relies on Karush–Kuhn–Tucker (KKT) conditions and a generalized eigen-decomposition method to find the exact solution. In addition, we also develop an approximate, practical algorithm to find a good beamforming matrix when using M-ary detection schemes. Numerical results are presented to assess the performance of the proposed methods across various scenarios.
\end{abstract}

\renewcommand\IEEEkeywordsname{Keywords}

\IEEEpeerreviewmaketitle

\section{Introduction}
\footnote{
*Corresponding author

Email addresses: \textbf{nguynam4@oregonstate.edu} (Nam Nguyen), \textbf{vuonga2@oregonstate.edu} (An Vuong), \textbf{nguyent11@etsu.edu} (Thuan Nguyen), \textbf{thinhq@oregonstate.edu} (Thinh Nguyen)

A part of this work was presented at IEEE Vehicular Technology Conference 2024, Washington, D.C., United States \cite{Nam2024}.}

The security of wireless communication has been a prominent concern. A passive eavesdropper located within the coverage area of a wireless transmission can covertly acquire information about the transmitted signal, eliminating the possibility of being detected. Although encryption plays a crucial role in preserving data confidentiality, its implementation comes with substantial computational costs and challenges in the secure distribution and management of encryption keys. Despite these measures, a compelling need remains to enhance transmission security further and reduce the risk of signal interception \cite{Schneier1998}. To that end, recent advancements in physical layer security, such as beamforming and artificial noise injection, have been proposed as effective strategies for enhancing wireless security \cite{Mukherjee2014, Liu2017}. These approaches leverage the information-theoretic secrecy properties of physical communication channels, initially pioneered by Wyner in the wiretap channel \cite{Wyner1975}. In this context, the transmitter (Alice) wants to transmit confidential information to the authorized receiver (Bob) while protecting it from a potential eavesdropper (Eve). Wyner's work demonstrates the feasibility of establishing a reliable and secure communication channel in the presence of eavesdroppers, particularly when the eavesdropper's signal-to-noise ratio (SNR) is lower than that of the legitimate receiver. Later research extended this result to other types of channels besides the wiretap channel, such as non-degraded discrete memoryless broadcast channels \cite{Csiszar1978}, and also applied it to the standard Gaussian channel \cite{Leung1978}.

As high-capacity multiple-input-multiple-output (MIMO) communication systems have evolved, numerous studies have effectively characterized the secrecy capacity of such systems \cite{Goel2005, Khisti2006, Khisti2007, Li2007}. Secrecy capacity is the maximum transmission rate at which the eavesdropper cannot decipher any information \cite{Gopala2008}. In this paper, instead of secrecy capacity, the beamforming schemes are designed to minimize the symbol error probability (SEP) for an authorized user while guaranteeing that the likelihood of an eavesdropper correctly recovering symbols remains below a predefined threshold. Specifically, this paper is focused on finding an optimal beamforming vector for binary antipodal signal detection in multiple-input multiple-output (MIMO) Gaussian wiretap channels. Finding the optimal beamforming vector in this setting is challenging. Computationally efficient algorithms such as convex techniques cannot be applied to find the optimal solution. To that end, our proposed algorithm relies on Karush–Kuhn–Tucker (KKT) conditions and a generalized eigen-decomposition method to find the exact solution. In addition, we also develop an approximate, practical algorithm to find a good beamforming matrix when using M-ary detection schemes. Numerical results are presented to assess the performance of the proposed methods across various scenarios.

\textit{Paper Outline:} This paper is structured as follows. We begin by reviewing related work in Section \ref{RelatedWork}. Section \ref{MIMOCommunication} introduces the symbol error probability-based MIMO beamforming and MIMO Gaussian wiretap channel. We consider the SEP-based binary antipodal beamforming in Section \ref{BinaryAntipodalBeamformingProblem}. In Section \ref{M_AryBeamformingProblem}, we explore the SEP-based M-ary beamforming. Section \ref{NumericalResults} presents numerical results and discussions. Finally, we conclude the paper in Section \ref{Conclusion}.

\section {Related Work}
\label{RelatedWork}
Several efforts have been made to find closed-form solutions for the secrecy capacity of the MIMO Gaussian wiretap channel, as discussed in \cite{Liu2009, Oggier2011}. In this context, the secrecy capacity involves optimizing the transmit covariance matrix of the input signal, which is a non-convex optimization problem. This fundamental problem in studying physical layer security in MIMO systems has prompted extensive research over the past decade, utilizing various approaches \cite{Chakravarty2019}. While the optimality of Gaussian signaling and a general formula of the secrecy capacity is well established, closed-form solutions for the optimal transmit covariance matrix exist only for specific cases, leaving the general case unresolved \cite{Parada2005, Fakoorian2013}.

Various iterative and sub-optimal solutions have been proposed to address the secrecy capacity, primarily focusing on determining the transmit covariance matrix. For instance, alternating optimization and convex reformulation algorithms have been introduced in \cite{Li2013, Loyka2015, Nguyen2020, Dong2020, Mukherjee2021}. These methods transform the non-convex problem into a convex one and iteratively solve it using convex optimization. However, the complexity of these methods is high, and their solutions can be unstable in specific antenna configurations.

In addition, generalized singular value decomposition (GSVD)-based precoding, which decomposes the transmit channel into multiple parallel subchannels, offers a closed-form solution \cite{Khisti2010, Fakoorian2012}. Nevertheless, in certain antenna settings, such as when the desired receiver has only one antenna, this closed-form solution deviates significantly from the achievable capacity. A new covariance matrix parameterization was proposed for transmitters with two antennas, and its optimal closed-form solution was derived in \cite{Vaezi2017}. This approach was generalized to arbitrary numbers of antennas in \cite{Zhang2021}. However, the iterative and computationally intensive nature of the parameter determination process, particularly for systems with many transmit antennas, still needs to be improved.

Recently, deep learning (DL)-based precoding techniques have been proposed for secure communication over MIMO wiretap channels \cite{Zhang2019, Tuan2020, Pauls2022, Wang2022}. While achieving near-capacity secure rates efficiently, DL-based precoding requires substantial training time and computational resources, and its performance is sensitive to the quality and quantity of training data.

While secrecy capacity is a commonly used metric, its practical implementation and measurement present challenges in real-world scenarios where practical non-Gaussian codes are utilized. To tackle this, linear beamforming techniques have been proposed to exploit the transmit diversity by weighting the information stream \cite{Joham2005, Wiesel2006, Yu2007}. Mukherjee et al. \cite{Mukherjee2011} explore SINR-based beamforming for security in MIMO Gaussian wiretap channels, focusing on designing beamforming schemes for single-data-stream transmission, with the SINR at the legitimate receiver serving as a quality of service (QoS) metric. As a result, robust algorithms have been developed to minimize the authorized receiver's required transmit power to attain the desired QoS, even when channel state information errors are present.

Our work introduces optimization frameworks that provide practical solutions for enhancing wireless security in the MIMO Gaussian wiretap channel. By focusing on optimizing the symbol error rate rather than the signal-to-interference-plus-noise ratio, we aim to design efficient algorithms that minimize the symbol error probability for legitimate users while ensuring that the probability of an eavesdropper correctly recovering symbols remains below a predetermined threshold. To the best of our knowledge, this approach has not been explored in the context of the MIMO Gaussian wiretap channel. 

\textit{Paper Notation:} $\mathbb{R}^{M}$ is the set of $M$-dimensional real vectors. $\mathbb{R}^{M\times N}$ is the set of $M \times N$ real matrices. $\mathbb{C}^{M}$ is the set of $M$-dimensional complex vectors. $\mathbb{C}^{M\times N}$ is the set of $M \times N$ complex matrices. Bold lower and upper case letters express vectors and matrices, respectively. Non-bold letters denote scalars. The operators $(.)^T$ and $(.)^H$ indicate transpose and complex conjugate transpose, respectively. $\textnormal{vec}(.)$ is the vectorization operator. $\textnormal{Tr}(.)$ is the trace operator. $|.|$ is the determinant operator. $\left \|. \right \|_2$ and $\left \|. \right \|_F$ are the Euclidean norm and Frobenius norm, respectively. $\Re [.]$ and $\Im [.]$ are the real and imaginary operators, respectively. $\mathbf{X} \succeq 0 $ means that $\mathbf{X}$ is symmetric positive semidefintie.

\section{MIMO beamforming in Gaussian wiretap channel}
\label{MIMOCommunication}
In this section, we characterize symbol error probability-based MIMO beamforming for both binary and M-ary signal detection and present the MIMO Gaussian wiretap channel.
\subsection{Symbol Error Probability}
Consider a MIMO communication system with $N$ transmit antennas and $K$ receive antennas. The MIMO received signal $\mathbf{y}$ can be expressed as:
\begin{equation}
\label{SystemModel}
\mathbf{y}=\mathbf{H}\mathbf{W}\mathbf{s}+\mathbf{n},
\end{equation}
where $\mathbf{H} \in \mathbb{C}^{K\times N}$ is channel matrix between transmitter and receiver as
\begin{equation}
\label{ChannelModel}
\mathbf{H} = [h_{ij}]_{i=1,...,K; j=1,...,N} = \begin{bmatrix}
\mathbf{h}_1\\ 
...\\ 
\mathbf{h}_K
\end{bmatrix},
\end{equation}
$\mathbf{W} \in \mathbb{C}^{N\times L}$ is the beamforming matrix. $L$ is the number of information symbols and $L$ does not exceed $\min\{N,K\}$. $\mathbf{s} = [s_1,...,s_L]^T$ is the information intended for the receiver, $\mathbf{n} \in \mathbb{C}^{K} \sim \mathcal{C}\mathcal{N} (0, N_0 \mathbf{I})$ is the complex Gaussian noise.  
For analysis, we assume  $\mathbf{H}$ is known, i.e., the full channel state information is available at both the transmitter and receiver side \cite{Sung2009, Stankovic2008, Hochwald2005}.

{\bf Binary Signal Detection.}
In this section, binary signals are sent, i.e.,  $\mathbf{s} \in \left \{ \mathbf{s}_0, \mathbf{s}_1\right \}$, where $\mathbf{s}_0 = [a_{0_1}, ..., a_{0L}]^T$ is an arbitrary complex $L$-dimensional vector. 

We assume the prior probabilities of the two signals $\mathbf{s}_0$ and $\mathbf{s}_1$ to be equal. Therefore, the error probabilities are the same whether the symbol vector $\mathbf{s}_0$ or $\mathbf{s}_1$ was transmitted and can be derived as \cite{Gallager2008}
\begin{eqnarray}
P_e^{Binary} &=& P(\hat{\mathbf{s}}=\mathbf{s}_1|\mathbf{s}=\mathbf{s}_0) = P(\hat{\mathbf{s}}=\mathbf{s}_0|\mathbf{s}=\mathbf{s}_1), \\
&=& Q\left ( \frac{\parallel \mathbf{H} \mathbf{W} (\mathbf{s}_0-\mathbf{s}_1) \parallel_2 }{2\sqrt{N_0/2}} \right ), \label{SymbolErrorProbabilityFormualtion}
\end{eqnarray}
where $\hat{\mathbf{s}}$ is the detected signal of $\mathbf{s}$ and $Q(x)=\frac{1}{\sqrt{2\pi}}\int_{x}^{\infty }e^{-\frac{t^2}{2}}dt$ is the Q-function.

{\bf $\mathbf{M}$-ary Signal Detection.}
We consider the set of complex vector  $\mathbf{s} \in \left \{ \mathbf{s}_1, ..., \mathbf{s}_M\right \}$, where each $\mathbf{s}_i = [a_{i1}, ..., a_{iL}]^T$ represents an arbitrary $L$-dimensional vector and $i \in \{1, ..., M\}$ indexes the vectors in the set.

According to the union bound in probability theory, the probability that at least one of a countable or finite set of events occurs is less than or equal to the sum of the probabilities of the individual events \cite{Boole1847, Lopes2023}. Therefore,
\begin{equation}
\textnormal{P}\left (\bigcup_{i}{B_i} \right ) \leq \sum_{i} \textnormal{P} (B_i),
\end{equation}				
		
Following the above formula, let $A_{ij}$ denote the event that $\parallel \hat{\mathbf{s}} - \mathbf{s}_j \parallel < \parallel \hat{\mathbf{s}} - \mathbf{s}_i \parallel$ $\forall i \neq j$ and $i, j \in \{1, ..., M\}$. The error probability of transmitting the symbol vector $\mathbf{s}_i$ is then expressed as
\begin{equation}
\textnormal{P}(\hat{\mathbf{s}} \neq \mathbf{s}_i|\mathbf{s}=\mathbf{s}_i)=\textnormal{P}(\bigcup_{j=1, j\neq i}^{M}A_{ij})\leq \sum_{j=1, j\neq i}^M \textnormal{P}(A_{ij}),
\end{equation}
where $\textnormal{P}(A_{ij})=Q\left ( \frac{\parallel \mathbf{H} \mathbf{W} (\mathbf{s}_i-\mathbf{s}_j) \parallel_2 }{2\sqrt{N_0/2}} \right )$.
		
Therefore,
\begin{equation}
\textnormal{P}(\hat{\mathbf{s}} \neq \mathbf{s}_i|\mathbf{s}=\mathbf{s}_i)\leq \sum_{j=1, j \neq i}^{M}Q\left ( \frac{\parallel \mathbf{H} \mathbf{W} (\mathbf{s}_i - \mathbf{s}_j)\parallel_2 }{2\sqrt{N_0/2}} \right ),
\end{equation} 
		
In summary, the union bound for symbol error probability is formulated as follows:
\begin{eqnarray}
P_e^{M-ary} &=& \sum_{i=1}^M \textnormal{P}(\hat{\mathbf{s}} \neq \mathbf{s}_i,\mathbf{s}=\mathbf{s}_i), \label{SymbolErrorProbabilityFormualtion}\\
& = & \sum_{i=1}^M\textnormal{P}(\hat{\mathbf{s}} \neq \mathbf{s}_i|\mathbf{s}=\mathbf{s}_i)P(\mathbf{s}=\mathbf{s}_i), \\
& \leq & \frac{1}{M}\sum_{i=1}^{M} \sum_{j=1, j \neq i}^{M} Q\left ( \frac{\parallel \mathbf{H}\mathbf{W} (\mathbf{s}_i-\mathbf{s}_j) \parallel_2 }{2\sqrt{N_0/2}} \right ) \nonumber\\
\label{UnionBoundSymbolErrorProbabilityFormualtion}
\end{eqnarray}
where $P(\mathbf{s}=\mathbf{s}_i)=\frac{1}{M}$ is the probability that the symbol vector $\mathbf{s}_i$ is transmitted.	
			
\subsection{MIMO Gaussian Wiretap Channel}
\begin{figure}
\begin{center}
\includegraphics[width=0.34\textwidth]{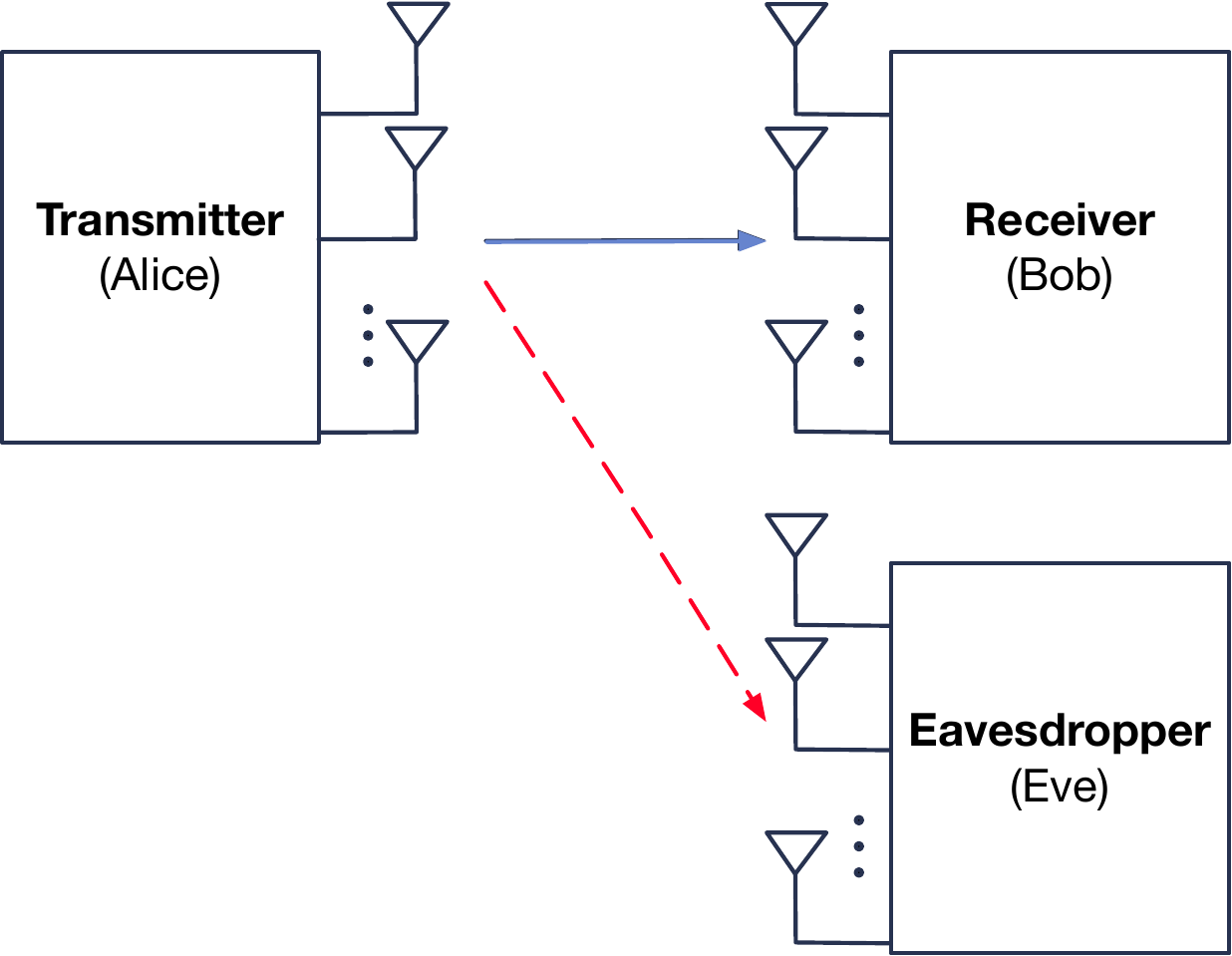}
\caption{MIMO Gaussian wiretap channel.}	
\label{MIMOwirtapChannel}	        
\end{center}	
\end{figure}
We consider a MIMO wiretap channel described in Figure \ref{MIMOwirtapChannel}. The number of antennas at the transmitter (Alice), the legitimate receiver (Bob), and the eavesdropper (Eve) are $N$, $K_B$, and $K_E$, respectively. We call the channel between Alice and Bob the main channel ($\mathbf{H}_B \in \mathbb{C}^{K_B \times N}$) while the channel between Alice and Eve is the eavesdropper channel ($\mathbf{H}_E \in \mathbb{C}^{K_E \times N}$). The scenario is that Alice sends her message to Bob while Eve overhears the information conveyed from Alice to Bob without interfering with the main channel \cite{Wyner1975, Khisti2010, Oggier2011}. 
	
Following the system model in Equation (\ref{SystemModel}), the received signals at Bob and Eve are expressed as follows
\begin{align}
\mathbf{y}_B &= \mathbf{H}_B\mathbf{W}\mathbf{s}+\mathbf{n}_B, \\
\mathbf{y}_E &= \mathbf{H}_E\mathbf{W}\mathbf{s}+\mathbf{n}_E,
\end{align}
where $\mathbf{n}_B \sim \mathcal{C}\mathcal{N} (0, N_B \mathbf{I})$ and $\mathbf{n}_E \sim \mathcal{C}\mathcal{N} (0, N_E \mathbf{I})$ are the noise for the channels of Bob and Eve, respectively. Moreover, the noise is independent of channel realization. The input signal is subjected to a power constraint $P$ such that $\textnormal{Tr} (\mathbf{W}\mathbf{W}^H) \leq P$ \cite{Christensen2008, Yang2013}.

\section{SEP-based Binary Antipodal Beamforming}
\label{BinaryAntipodalBeamformingProblem}
\subsection{Problem Formulation}
\label{BeamformingProblem1}
In this work, we focus on enhancing reliability and signal-to-noise ratio rather than maximizing the bit rate. Consequently, we investigate a special case of binary antipodal signals, where the information symbols are $a$ or $-a$ with $a \in \mathbb{C}$. The binary antipodal signals can be expressed as $\mathbf{s} = \mathbf{w}a$ or $\mathbf{s} = \mathbf{w}(-a)$, where $\mathbf{w} \in \mathbb{C}^{N}$ is the beamforming vector. Thus, this section is dedicated to analyzing a single data stream using beamforming.

Based on (\ref{SymbolErrorProbabilityFormualtion}), the error probabilities of Bob and Eve are expressed as follows:
\begin{align}
P_e^B &= Q\left ( \frac{\parallel  \mathbf{H}_B \mathbf{w}a\parallel_2}{\sqrt{N_B/2}} \right ), \\
P_e^E &= Q\left ( \frac{\parallel  \mathbf{H}_E \mathbf{w}a\parallel_2}{\sqrt{N_E/2}} \right ),
\end{align}

We concentrate on cases where Alice transmits a single data stream to Bob. Our goal is to minimize Bob's error probability while ensuring that Eve's error probability exceeds a predetermined threshold, all while meeting the power constraint. Consequently, we formulate the following optimization problem:
\begin{mini!}|s|[2] 
{\mathbf{w}} 
{Q\left ( \frac{\parallel  \mathbf{H}_B \mathbf{w}a\parallel_2 }{\sqrt{N_B/2}} \right )} 
{\label{GeneralProbBemforming}} 
{} 
\addConstraint{Q\left ( \frac{\parallel  \mathbf{H}_E \mathbf{w}a\parallel_2 }{\sqrt{N_E/2}} \right )}{\geq  D \label{},} 
\addConstraint{\left \| \mathbf{w}  \right \|_2^2}{\leq P, \label{}} 
\end{mini!}
where $\mathbf{H}_B \in \mathbb{C}^{K_B\times N}$, $\mathbf{H}_E \in \mathbb{C}^{K_E\times N}$, $\mathbf{w} \in \mathbb{C}^{N}$, $a \in \mathbb{C}$, $D \in [0,0.5]$, and $P > 0$ is the transmitted signal power.

The concept of a secrecy codebook plays a crucial role in designing secure communication strategies. In essence, beamforming techniques act as a practical application of this concept, where secret messages are mapped to specific beamformers with designated transmit powers to ensure secure transmission.

Let $\bar{\mathbf{w}} = \frac{\mathbf{w}}{\sqrt{P}}$, the problem (\ref{GeneralProbBemforming}) can be rewritten as
\begin{mini!}|s|[2] 
{\bar{\mathbf{w}}} 
{Q\left ( \sqrt{\frac{2P}{N_B}} \parallel  \mathbf{H}_B \bar{\mathbf{w}}a\parallel_2 \right ) \label{GeneralProbBemforming_Normalize_Objective}} 
{\label{GeneralProbBemforming_Normalize}} 
{} 
\addConstraint{Q\left ( \sqrt{\frac{2P}{N_E}} \parallel  \mathbf{H}_E \bar{\mathbf{w}}a\parallel_2 \right )}{\geq  D \label{},} 
\addConstraint{\left \| \bar{\mathbf{w}} \right \|_2^2}{\leq 1, \label{}} 
\end{mini!}
 
Due to the Q-function being a monotonically decreasing function, an equivalent problem of  (\ref{GeneralProbBemforming_Normalize}) is: 
\begin{maxi!}|s|[2] 
{\bar{\mathbf{w}}} 
{\parallel  \mathbf{H}_{B} \bar{\mathbf{w}}\parallel_2^2} 
{\label{GeneralProbBemforming_Normalize_Equivalent}} 
{} 
\addConstraint{\parallel  \mathbf{H}_{E} \bar{\mathbf{w}}\parallel_2^2}{\leq  \left (\frac{\sqrt{N_E/(2P)}Q^{-1}(D)}{|a|} \right )^2, \label{}} 
\addConstraint{\left \| \bar{\mathbf{w}}  \right \|_2^2}{\leq 1 , \label{}} 
\end{maxi!}
The objective of the problem (\ref{GeneralProbBemforming_Normalize_Equivalent}) is a convex function, where the feasible space of this problem is compact and convex. Therefore, an optimal solution must lie at extreme points \cite{Tuy1985, Hoffman1981}. Finding these extreme points is challenging, so KKT conditions are used to develop an efficient algorithm.
\begin{figure*}
\begin{eqnarray}
\mathcal{L}(\bar{\mathbf{w}},\lambda_1,\lambda_2) &=& -\parallel  \mathbf{H}_{B} \bar{\mathbf{w}} \parallel_2^2 + \lambda_1 \left ( \parallel \mathbf{H}_{E} \bar{\mathbf{w}} \parallel_2^2 - \left ( \frac{\sqrt{N_E/(2P)}Q^{-1}(D)}{|a|} \right )^2 \right ) + \lambda_2 (\left \| \bar{\mathbf{w}}  \right \|_2^2 -1), \\
& = & -\parallel \mathbf{H}_{B} \bar{\mathbf{w}} \parallel_2^2 + \lambda_1 \parallel \mathbf{H}_{E} \bar{\mathbf{w}} \parallel_2^2 + \lambda_2 \left \| \bar{\mathbf{w}}  \right \|_2^2 - \lambda_2 - \lambda_1 \left ( \frac{\sqrt{N_E/(2P)}Q^{-1}(D)}{|a|} \right )^2, \\
&  = & - \bar{\mathbf{w}}^H \mathbf{H}_{B}^H \mathbf{H}_{B} \bar{\mathbf{w}} + \lambda_1 \bar{\mathbf{w}}^H \mathbf{H}_{E}^H \mathbf{H}_{E} \bar{\mathbf{w}} + \lambda_2 \left \| \bar{\mathbf{w}}  \right \|_2^2 - \lambda_2 - \lambda_1 \left ( \frac{\sqrt{N_E/(2P)}Q^{-1}(D)}{|a|} \right )^2, \\
& = & \bar{\mathbf{w}}^H (\lambda_1 \mathbf{H}_{E}^H \mathbf{H}_{E} - \mathbf{H}_{B}^H \mathbf{H}_{B} ) \bar{\mathbf{w}} + \lambda_2 \left \| \bar{\mathbf{w}}  \right \|_2^2 - \lambda_2 - \lambda_1 \left ( \frac{\sqrt{N_E/(2P)}Q^{-1}(D)}{|a|} \right )^2.
\label{LagrangeEq}
\end{eqnarray}
\end{figure*}

\subsection{KKT Conditions}
\label{sec:KKT}
The Lagrangian of the objective function is shown in Equation (\ref{LagrangeEq}). The gradient of the Lagrangian can be derived as follows 
\begin{equation}
\nabla_{\bar{\mathbf{w}}} \mathcal{L}(\bar{\mathbf{w}},\lambda_1,\lambda_2) = 2  (\lambda_1 \mathbf{H}_{E}^H \mathbf{H}_{E} - \mathbf{H}_{B}^H \mathbf{H}_{B} ) \bar{\mathbf{w}} + 2 \lambda_2 \bar{\mathbf{w}}.
\end{equation}
	
Therefore, the KKT conditions of the problem (\ref{GeneralProbBemforming_Normalize_Equivalent}) are:
\begin{subequations}
\label{KKT_conditions}
\begin{align}
& \|\mathbf{H}_{E} \bar{\mathbf{w}}^*\|_2^2 - \left(\frac{\sqrt{N_E/(2P)}Q^{-1}(D)}{|a|}\right)^2 \leq 0, \label{eq:KKT1} \\
& \|\bar{\mathbf{w}}^*\|_2^2 - 1 \leq 0, \label{eq:KKT2} \\
& \lambda_1^*, \lambda_2^* \geq 0, \label{eq:KKT3} \\
& \lambda_1^* \left(\|\mathbf{H}_{E} \bar{\mathbf{w}}^*\|_2^2 - \left(\frac{\sqrt{N_E/(2P)}Q^{-1}(D)}{|a|}\right)^2\right) = 0, \label{eq:KKT4} \\
& \lambda_2^* \left(\|\bar{\mathbf{w}}^*\|_2^2 - 1\right) = 0, \label{eq:KKT5} \\ 
& (\lambda_1^* \mathbf{H}_{E}^H \mathbf{H}_{E} - \mathbf{H}_{B}^H \mathbf{H}_{B}) \bar{\mathbf{w}}^* + \lambda_2^* \bar{\mathbf{w}}^* = \mathbf{0}. \label{eq:KKT6}
\end{align}
\end{subequations}

An optimal solution must satisfy the KKT conditions. There are $4$ cases corresponding to the possibly optimal values of $\lambda_1^*$ and $\lambda_2^*$.
\begin{itemize}
    \item \underline{Case 1:} For $\lambda_1^* = 0$ and $\lambda_2^* = 0$,
    \begin{subequations}
    \label{Case1Constraint}
    \begin{align}
    & \|\mathbf{H}_{E} \bar{\mathbf{w}}^*\|_2^2 - \left( \frac{\sqrt{N_E/(2P)}Q^{-1}(D)}{|a|}\right)^2 < 0, \label{Case1Constraint1} \\
    & \|\bar{\mathbf{w}}^*\|_2^2 - 1 < 0, \label{Case1Constraint2} \\
    & (\mathbf{H}_{B}^H \mathbf{H}_{B}) \bar{\mathbf{w}}^* = \mathbf{0}. \label{Case1Constraint3}
    \end{align}
    \end{subequations}
	
    \item \underline{Case 2:} For $\lambda_1^* = 0$ and $\lambda_2^* > 0$,
    \begin{subequations}
    \label{Case2Constraint}
    \begin{align}
    & \|\mathbf{H}_{E} \bar{\mathbf{w}}^*\|_2^2 - \left( \frac{\sqrt{N_E/(2P)}Q^{-1}(D)}{|a|}\right)^2 < 0, \label{Case2Constraint1} \\
    & \|\bar{\mathbf{w}}^*\|_2^2 - 1 = 0, \label{Case2Constraint2} \\
    & \mathbf{H}_{B}^H \mathbf{H}_{B}  \bar{\mathbf{w}}^* = \lambda_2^* \bar{\mathbf{w}}^*. \label{Case2Constraint3}
    \end{align}
    \end{subequations}

    \item \underline{Case 3:} For $\lambda_1^* > 0$ and $\lambda_2^* = 0$,
    \begin{subequations}
    \label{Case3Constraint}
    \begin{align}
    & \|\mathbf{H}_{E} \bar{\mathbf{w}}^*\|_2^2 - \left( \frac{\sqrt{N_E/(2P)}Q^{-1}(D)}{|a|}\right)^2 = 0, \label{Case3Constraint1} \\
    & \|\bar{\mathbf{w}}^*\|_2^2 - 1 < 0, \label{Case3Constraint2} \\
    & \mathbf{H}_{B}^H \mathbf{H}_{B} \bar{\mathbf{w}}^* = \lambda_1^* \mathbf{H}_{E}^H \mathbf{H}_{E}\bar{\mathbf{w}}^*. \label{Case3Constraint3}
    \end{align}
    \end{subequations}

    \item \underline{Case 4:} For $\lambda_1^* > 0$ and $\lambda_2^* > 0$,
    \begin{subequations}
    \label{Case4Constraint}
    \begin{align}
    & \|\mathbf{H}_{E} \bar{\mathbf{w}}^*\|_2^2 - \left( \frac{\sqrt{N_E/(2P)}Q^{-1}(D)}{|a|}\right)^2 = 0, \label{Case4Constraint1} \\
    & \|\bar{\mathbf{w}}^*\|_2^2 - 1 = 0, \label{Case4Constraint2} \\
    & (\mathbf{H}_{B}^H \mathbf{H}_{B} - \lambda_1^* \mathbf{H}_{E}^H \mathbf{H}_{E}) \bar{\mathbf{w}}^* = \lambda_2^* \bar{\mathbf{w}}^*. \label{Case4Constraint3}
    \end{align}
    \end{subequations}
\end{itemize}
\begin{figure*}
\centering
\begin{subfigure}{0.32\textwidth}
\centering
\includegraphics[width=\linewidth]{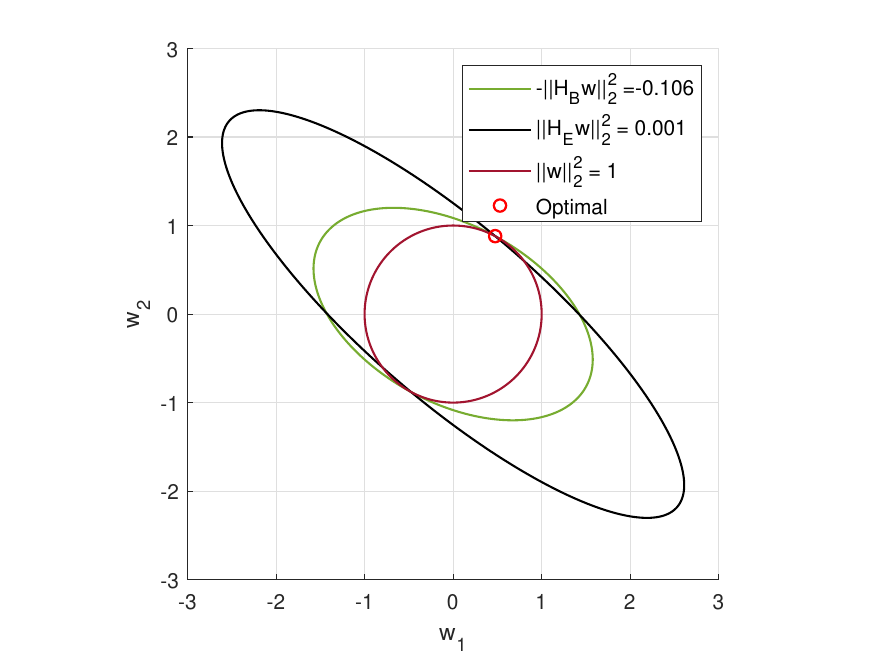}
\caption{Setup 1}
\label{Setup1}
\end{subfigure}
\hfill
\begin{subfigure}{0.32\textwidth}
\centering
\includegraphics[width=\linewidth]{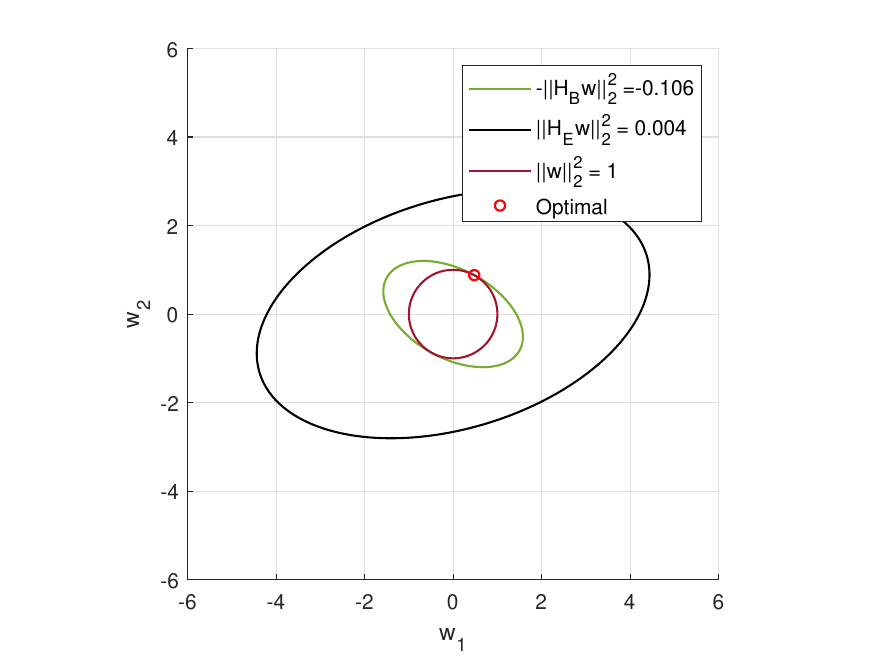}
\caption{Setup 2}
\label{Setup2}
\end{subfigure}
\hfill
\begin{subfigure}{0.32\textwidth}
\centering
\includegraphics[width=\linewidth]{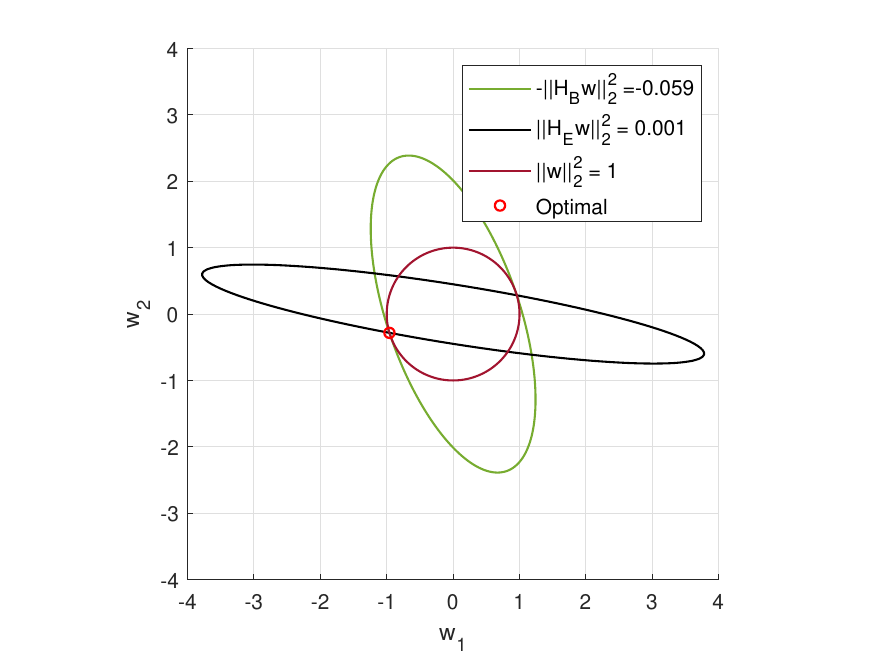}
\caption{Setup 3}
\label{Setup3}
\end{subfigure}
\caption{Illustration of the optimal case of (a) setup 1, (b) setup 2, and setup 3 in $\mathbb{R}^2$.}
\end{figure*}

\textbf{The upper bound of $\lambda_1^*$ in Case 4.} From (\ref{Case4Constraint3}), we have:
\begin{equation*}
(\mathbf{H}_{B}^H \mathbf{H}_{B} - \lambda_1^* \mathbf{H}_{E}^H \mathbf{H}_{E}) \bar{\mathbf{w}}^* = \lambda_2^* \bar{\mathbf{w}}^*,
\end{equation*}
where $\lambda_1^*, \lambda_2^* > 0$. 

For $\lambda_2 > 0$ to be true, the matrix $\mathbf{H}_{B}^H \mathbf{H}_{B} - \lambda_1^* \mathbf{H}_{E}^H \mathbf{H}_{E}$ must be positive definite. This requires all eigenvalues of this matrix to be positive. The positive definiteness condition can be expressed as: 
\begin{equation}
\mathbf{H}_{B}^H \mathbf{H}_{B} - \lambda_1^* \mathbf{H}_{E}^H \mathbf{H}_{E} \succ 0,
\end{equation}

Consider any nonzero vector $\mathbf{u}$, we have:
\begin{eqnarray}
\mathbf{u}^H (\mathbf{H}_{B}^H \mathbf{H}_{B} - \lambda_1^*  \mathbf{H}_{E}^H \mathbf{H}_{E}) \mathbf{u} &>& 0, \\ \nonumber \\
 \mathbf{u}^H \mathbf{H}_{B}^H \mathbf{H}_{B} \mathbf{u} - \lambda_1^* \mathbf{u}^H \mathbf{H}_{E}^H \mathbf{H}_{E} \mathbf{u} &>& 0, \\ \nonumber \\
\lambda_1^* < \frac{\mathbf{u}^H \mathbf{H}_{B}^H \mathbf{H}_{B} \mathbf{u}}{\mathbf{u}^H \mathbf{H}_{E}^H \mathbf{H}_{E} \mathbf{u}},
\end{eqnarray}

From the Rayleigh-Ritz theorem, it holds that
\begin{equation}
\lambda_{\text{min}}(\mathbf{H}_{B}^H \mathbf{H}_{B})\mathbf{u} \leq \mathbf{u}^H \mathbf{H}_{B}^H \mathbf{H}_{B} \mathbf{u} \leq \lambda_{\text{max}}(\mathbf{H}_{B}^H \mathbf{H}_{B})\mathbf{u},
\end{equation}
\begin{equation}
\lambda_{\text{min}}(\mathbf{H}_{E}^H \mathbf{H}_{E})\mathbf{u} \leq \mathbf{u}^H \mathbf{H}_{E}^H \mathbf{H}_{E} \mathbf{u} \leq \lambda_{\text{max}}(\mathbf{H}_{E}^H \mathbf{H}_{E})\mathbf{u},
\end{equation}
where $\lambda_{\text{min}}(\mathbf{X})$ and $\lambda_{\text{max}}(\mathbf{X})$ denote the minimum and maximum eigenvalues of $\mathbf{X}$, respectively.

Hence, 
\begin{eqnarray}
\lambda_1^* < \frac{\lambda_{\text{max}}(\mathbf{H}_{B}^H \mathbf{H}_{B})}{\lambda_{\text{min}}(\mathbf{H}_{E}^H \mathbf{H}_{E})},
\end{eqnarray}

Therefore, the range of $\lambda_1^*$ is expressed as follows:
\[ 0 < \lambda_1^* < \frac{\lambda_{\text{max}}(\mathbf{H}_{B}^H \mathbf{H}_{B})}{\lambda_{\text{min}}(\mathbf{H}_{E}^H \mathbf{H}_{E})}. \]

\subsection{SEP--Antipodal Beamforming Algorithm}
\begin{algorithm}
\caption{SEP--Antipodal Beamforming.}
\label{AlgorithmBeamforming}
\begin{algorithmic}[1]
\renewcommand{\algorithmicrequire}{\textbf{Input:}}
\renewcommand{\algorithmicensure}{\textbf{Output:}}
\REQUIRE $\mathbf{H}_{B}$, $\mathbf{H}_{E}$, $N_B$, $N_E$, $D$, $P$, $a$
\ENSURE $\bar{\mathbf{w}}^*$
\STATE \textbf{Initialization:} Determine the optimal $\bar{\mathbf{w}}^*$ considering Cases 2, 3, and 4.\\
\underline{Case 2:}
\STATE Find the eigenvector set $\mathbb{E}_2$ such that $\mathbf{H}_{B}^H \mathbf{H}_{B}  \bar{\mathbf{w}} = \lambda_2 \bar{\mathbf{w}}$;
\IF {$\lambda_2 > 0$ and $\bar{\mathbf{w}}_2 \in \mathbb{E}_2$ minimizes (\ref{GeneralProbBemforming_Normalize_Objective}) and satisfies (\ref{Case2Constraint1}), (\ref{Case2Constraint2})}
	\STATE Compute $P_{e_2}^B = Q\left ( \sqrt{2P/N_B} \parallel  \mathbf{H}_{B} \bar{\mathbf{w}}_2 a \parallel_2 \right )$;
\ELSE
	\STATE Set $P_{e_2}^B = +\infty$;
\ENDIF \\
\underline{Case 3:}
\STATE Find the generalized eigenvector set $\mathbb{E}_3$ such that $\mathbf{H}_{B}^H \mathbf{H}_{B} \bar{\mathbf{w}} = \lambda_1 \mathbf{H}_{E}^H \mathbf{H}_{E} \bar{\mathbf{w}}$;
\IF {$\lambda_1 > 0$ and $\bar{\mathbf{w}}_3 \in \mathbb{E}_3$ minimizes (\ref{GeneralProbBemforming_Normalize_Objective}) and satisfies (\ref{Case3Constraint1}), (\ref{Case3Constraint2})}
	\STATE Compute $P_{e_3}^B = Q\left ( \sqrt{2P/N_B} \parallel  \mathbf{H}_{B} \bar{\mathbf{w}}_3 a \parallel_2 \right )$;
\ELSE
	\STATE Set $P_{e_3}^B = +\infty$;
\ENDIF \\
\underline{Case 4:}
\STATE Perform an exhaustive search with resolution $\epsilon$ over all values of $\lambda_1 \in \left (  0, \frac{\lambda_{\text{min}}(\mathbf{H}_{B}^H \mathbf{H}_{B})}{\lambda_{\text{max}}(\mathbf{H}_{E}^H \mathbf{H}_{E})} \right )$.
\FOR {each value of $\lambda_1$}
    \STATE Find the eigenvector set $\mathbb{E}_4$ such that $(\mathbf{H}_{B}^H \mathbf{H}_{B} - \lambda_1 \mathbf{H}_{E}^H \mathbf{H}_{E}) \bar{\mathbf{w}} = \lambda_2 \bar{\mathbf{w}}$;
    \IF {$\lambda_2 > 0$ and $\bar{\mathbf{w}}_4 \in \mathbb{E}_4$ minimizes (\ref{GeneralProbBemforming_Normalize_Objective}) and satisfies (\ref{Case4Constraint1}), (\ref{Case4Constraint2})}
        \STATE Compute $P_{e_4}^B = Q\left ( \sqrt{2P/N_B} \parallel  \mathbf{H}_{B} \bar{\mathbf{w}}_4 a \parallel_2 \right )$;
    \ELSE
        \STATE Set $P_{e_4}^B = +\infty$;
    \ENDIF
\ENDFOR
\STATE Select $\bar{\mathbf{w}}^* = \bar{\mathbf{w}}_{i^*}$ where $i^* = \min_i P_{e_i}^B = \min_i(P_{e_2}^B, P_{e_3}^B, P_{e_4}^B)$ and $\bar{\mathbf{w}}_{i} \in \left \{ \bar{\mathbf{w}}_2, \bar{\mathbf{w}}_3, \bar{\mathbf{w}}_4 \right \}$;
\end{algorithmic}
\end{algorithm}

The SEP-antipodal beamforming algorithm, detailed in Algorithm \ref{AlgorithmBeamforming}, iteratively explores Cases 2, 3, and 4 to determine the optimal beamforming vector, denoted as \(\bar{\mathbf{w}}^*\). This algorithm aims to minimize the symbol error probability for a legitimate user, with a focus on solutions located at the extreme points of the feasible region. Case 1 is excluded, as it yields solutions within the interior of the constraint set, which are suboptimal for this objective.

In Cases 2 and 3, the optimal beamforming vectors can be found efficiently by solving generalized eigenvalue problems. Specifically:
\begin{itemize}
    \item \textbf{Case 2} finds the eigenvector set \(\mathbb{E}_2\) that satisfies \(\mathbf{H}_{B}^H \mathbf{H}_{B} \bar{\mathbf{w}} = \lambda_2 \bar{\mathbf{w}}\) with \(\lambda_2 > 0\). If an eigenvector \(\bar{\mathbf{w}}_2\) from this set minimizes the objective function and satisfies the predefined constraints, it computes the probability \(P_{e_2}^B\).
    \item \textbf{Case 3} involves the generalized eigenvalue equation \(\mathbf{H}_{B}^H \mathbf{H}_{B} \bar{\mathbf{w}} = \lambda_1 \mathbf{H}_{E}^H \mathbf{H}_{E} \bar{\mathbf{w}}\). Here, \(\lambda_1 > 0\), and if an eigenvector \(\bar{\mathbf{w}}_3\) minimizes the objective function and satisfies the predefined constraints, \(P_{e_3}^B\) is computed accordingly.
\end{itemize}

\textbf{Case 4} requires an exhaustive search over \(\lambda_1\) within the interval \(\left(0, \frac{\lambda_{\text{max}}(\mathbf{H}_{B}^H \mathbf{H}_{B})}{\lambda_{\text{min}}(\mathbf{H}_{E}^H \mathbf{H}_{E})}\right)\) with a resolution of \(\epsilon\). For each \(\lambda_1\) value, the eigenvector set \(\mathbb{E}_4\) that satisfies \((\mathbf{H}_{B}^H \mathbf{H}_{B} - \lambda_1 \mathbf{H}_{E}^H \mathbf{H}_{E}) \bar{\mathbf{w}} = \lambda_2 \bar{\mathbf{w}}\) is determined with \(\lambda_2 > 0\). If a feasible eigenvector \(\bar{\mathbf{w}}_4\) satisfies the predefined constraints and minimizes the objective function, \(P_{e_4}^B\) is computed.

Finally, the algorithm selects \(\bar{\mathbf{w}}^*\) as the eigenvector corresponding to the minimum error probability among the computed probabilities \(P_{e_2}^B\), \(P_{e_3}^B\), and \(P_{e_4}^B\), ensuring that \(\bar{\mathbf{w}}^*\) optimally minimizes the error probability across all cases considered.

\textbf{Time complexity.}
In terms of complexity, our proposed algorithm involves finding all eigenvalues for a dense $N\times N$ matrix, which is of $O(N^{3})$. Additionally, in case 4, we need to search through $\lambda_{1}$, adding to the complexity. Therefore, the overall complexity of our algorithm can be expressed as $O(V N^{3})$, where $V$ represents the size of the search space.

\subsection{Case Analysis}
We present the analytical results based on different cases in Section \ref{sec:KKT}.  To simplify the analysis, the real-valued deterministic channel matrices are used. Figure 2 shows the objective function contours and constraint boundaries of three different setups. Note that in all three figures, the symmetry of the quadratic objective function determines two symmetric optimal points.  Also, all the optimal points are the extreme points as predicted for an optimization problem with a concave objective the feasible space is compact and convex set.

{ \bf Setup 1.}
\label{SetupReal1}
We consider the system parameters as $\mathbf{H}_B = \begin{bmatrix}
 0.21 & 0.011 \\ 
 0.09 & 0.3
\end{bmatrix}$, $\mathbf{H}_E = \begin{bmatrix}
 0.01 & 0.02 \\ 
 0.017 & 0.01
\end{bmatrix}$, $D = 0.346$, $P = 1$ W, $N_B = N_E = 0.01$ W, $N = K_B = K_E = M = 2, a \in \{-1,1\}$. For case 2, $P_{e_2}^B = 0.0035$, $P_{e_2}^E = 0.4427$ (i.e., the error probability of Eve), $\bar{\mathbf{w}}_2 = \begin{bmatrix}
-0.8784 & 
0.4779
\end{bmatrix}^T$, and $\lambda_2 = 0.0363$. For case 3, there exists no $\bar{\mathbf{w}}_3$ that satisfies the constraint (\ref{Case3Constraint1}). For case 4, $P_{e_4}^B = 2.0542 \times 10^{-6}$, $P_{e_2}^E = 0.346$, $\bar{\mathbf{w}}_4 = \begin{bmatrix}
0.475& 
0.88
\end{bmatrix}^T$, $\lambda_1 = 1.5$, and $\lambda_2 = 0.0361$. Overall, case 4 is selected and presented in Figure \ref{Setup1}. Despite similar channel matrix directions for Bob and Eve in Figure \ref{Setup1}, Eve's error probability worsens due to her channel's high attenuation coefficient. Case 4 is rare in practice as optimal points must be among extreme points generated by specific constraints.
				
{ \bf Setup 2.}
\label{SetupReal2}
The same setting parameters as the setup 1 are considered, except $\mathbf{H}_B = \begin{bmatrix}
 0.21 & 0.011 \\ 
 0.09 & 0.3
\end{bmatrix}$, $\mathbf{H}_E = \begin{bmatrix}
 -0.01 & 0.02 \\ 
 0.01 & 0.01
\end{bmatrix}$, and $D = 0.2$. For case 2, $P_{e_2}^B = 2.0541\times 10^{-6}$, $P_{e_2}^E = 0.3960$, $\bar{\mathbf{w}}_2 = \begin{bmatrix}
0.4779&
0.8784
\end{bmatrix}^T$, and $\lambda_2 = 0.1061$. For case 3 and case 4, there exists no $\bar{\mathbf{w}}_3$ and $\bar{\mathbf{w}}_4$ that satisfies the constraint (\ref{Case3Constraint1}) and (\ref{Case4Constraint1}), respectively. Hence, we choose case 2 and describe in Figure \ref{Setup2}. In this scenario, Bob and Eve's channel matrices are highly orthogonal. Using the beamforming vector directs the signal mainly to Bob, reducing Bob's error probability but increasing Eve's.
										
{ \bf Setup 3.}
\label{SetupReal3}
We keep the same setting parameters as the setup 1, except $\mathbf{H}_B = \begin{bmatrix}
 0.21 & 0.015 \\ 
 0.1 & 0.12
\end{bmatrix}$, $\mathbf{H}_E = \begin{bmatrix}
 0.01 & 0.071 \\ 
 0.01 & 0.01
\end{bmatrix}$, and $D = 0.3246$. For case 2 and case 3, there exists no $\bar{\mathbf{w}}_2$ and $\bar{\mathbf{w}}_3$ that satisfies the constraint (\ref{Case2Constraint1}) and (\ref{Case3Constraint1}), respectively. For case 4, $P_{e_4}^B = 2.9105\times 10^{-4}$ with respect to $\lambda_1 = 1$ and $\lambda_2 = 0.0053$. $P_{e_4}^E = 0.3246$ and $\bar{\mathbf{w}}_4 = \begin{bmatrix}
-0.9592& 
-0.2828
\end{bmatrix}^T$. Accordingly, case 4 is selected and represented in Figure \ref{Setup3}.

\subsection{Alternative Problem Formulation}
\label{OtherProblemFormulation}
This section addresses a different problem when designing the optimal beamforming vector. We aim to maximize Eve's error probability while ensuring that Bob's error probability remains within a predetermined threshold while adhering to the power constraint. Consequently, we formulate the optimization problem as follows:
\begin{maxi!}|s|[2] 
{\mathbf{w}} 
{Q\left ( \frac{\parallel  \mathbf{H}_E \mathbf{w}a\parallel_2 }{\sqrt{N_E/2}} \right )} 
{} 
{} 
\addConstraint{Q\left ( \frac{\parallel  \mathbf{H}_B \mathbf{w}a\parallel_2 }{\sqrt{N_B/2}} \right )}{\leq   D \label{},} 
\addConstraint{\left \| \mathbf{w}  \right \|_2^2}{\leq P, \label{}} 
\end{maxi!}
where $\mathbf{H}_B \in \mathbb{C}^{K_B \times N}$, $\mathbf{H}_E \in \mathbb{C}^{K_E\times N}$, $\mathbf{w} \in \mathbb{C}^{N}$, $a \in \mathbb{C}$, $D \in [0,0.5]$, and $P>0$.

Since the Q-function is a strictly decreasing function, we can derive the equivalent problem as
\begin{mini!}|s|[2] 
{\mathbf{w}} 
{\left \| \mathbf{H}_E \mathbf{w}\right \|_2^2} 
{\label{SpeacialcaseProblem}} 
{} 
\addConstraint{\left \| \mathbf{H}_B \mathbf{w}\right \|_2^2}{\geq \left (  \frac{\sqrt{N_B/2}Q^{-1}(D)}{|a|}\right )^2,\label{}} 
\addConstraint{\left \| \mathbf{w}  \right \|_2^2}{\leq P, \label{}} 
\end{mini!}

We observe the fact that,
\begin{eqnarray}
\left \| \mathbf{H}_B \mathbf{w}\right \|_2^2 & = & \mathbf{w}^H \mathbf{H}_B^H \mathbf{H}_B \mathbf{w} = \textnormal{Tr}(\mathbf{w}^H \mathbf{H}_B^H \mathbf{H}_B \mathbf{w}), \\
& = & \textnormal{Tr}(\mathbf{H}_B^H \mathbf{H}_B \mathbf{w} \mathbf{w}^H),
\end{eqnarray}

Similarly, we have:
\begin{eqnarray}
\left \| \mathbf{H}_E \mathbf{w}\right \|_2^2 & = \mathbf{w}^H \mathbf{H}_E^H \mathbf{H}_E \mathbf{w} = \textnormal{Tr}(\mathbf{w}^H \mathbf{H}_E^H \mathbf{H}_E \mathbf{w}), \\
& = \textnormal{Tr}(\mathbf{H}_E^H \mathbf{H}_E \mathbf{w} \mathbf{w}^H),
\end{eqnarray}

Let $\mathbf{A} = \mathbf{w} \mathbf{w}^H$, we obtain the equivalent optimization problem of (\ref{SpeacialcaseProblem}) as follows:
\begin{mini!}|s|[2] 
{\mathbf{A}} 
{\textnormal{Tr}(\mathbf{H}_E^H \mathbf{H}_E \mathbf{A})} 
{\label{Specail_SD}} 
{} 
\addConstraint{\textnormal{Tr}(\mathbf{H}_B^H \mathbf{H}_B \mathbf{A})}{\geq \left (  \frac{\sqrt{N_B/2}Q^{-1}(D)}{|a|}\right )^2,\label{}} 
\addConstraint{\textnormal{Tr}(\mathbf{A})}{\leq P, \label{}} 
\addConstraint{\mathbf{A}}{\succeq 0, \label{}} 
\addConstraint{ \textnormal{rank}(\mathbf{A})}{= 1, \label{non-convexConstraint}} 
\end{mini!}

The only non-convex constraint is (\ref{non-convexConstraint}). Thus, we drop it to obtain the following semidefinite relaxation (SDR) version of (\ref{Specail_SD}) as proposed by Ma et al. \cite{Ma2002}: 
 \begin{mini!}|s|[2] 
{\mathbf{A}} 
{\textnormal{Tr}(\mathbf{H}_E^H \mathbf{H}_E \mathbf{A})} 
{\label{Specail_SDR_Relaxation}} 
{} 
\addConstraint{\textnormal{Tr}(\mathbf{H}_B^H \mathbf{H}_B \mathbf{A})}{\geq \left (  \frac{\sqrt{N_B/2}Q^{-1}(D)}{|a|}\right )^2,\label{}} 
\addConstraint{\textnormal{Tr}(\mathbf{A})}{\leq P, \label{}} 
\addConstraint{\mathbf{A}}{\succeq 0, \label{}} 
\end{mini!}
 
{\bf Time complexity.} The semidefinite relaxation problem above can be efficiently addressed using interior-point methods \cite{Boyd2004}. Specifically, the convex optimization toolbox CVX \cite{Grant2009} is utilized to solve the Problem (\ref{Specail_SDR_Relaxation}) in MATLAB. The computational complexity of these methods typically does not exceed $O((K + N^2)^{3.5})$, although practical observations often reveal a substantially lower complexity. However, it is important to note that the optimal solution to (\ref{Specail_SDR_Relaxation}) may not be of rank one. As a result, it becomes necessary to derive a feasible solution $\tilde{\mathbf{w}}$ for (\ref{SpeacialcaseProblem}) from the SDR solution $\mathbf{A}^*$.

{\bf Get $\tilde{\mathbf{w}}$ from $\mathbf{A}^*$.} We employ the randomization method from \cite{Sidiropoulos2006}. This involves performing the eigen-decomposition of $\mathbf{A}^*$ as $\mathbf{A}^* = \mathbf{U} \mathbf{\Sigma} \mathbf{U}^H$ and selecting $\mathbf{w}_l$ such that $\mathbf{w}_l = \mathbf{U} \mathbf{\Sigma}^{1/2} \mathbf{e}_l$. Here, the elements of $\mathbf{e}_l$ are independent random variables uniformly distributed on the unit circle in the complex plane, specifically $[\mathbf{e}_l]_i = e^{j \theta _{l, i}}$, where $\theta _{l, i}$ are independent and uniformly distributed in $[0, 2\pi)$. This guarantees that $\mathbf{w}_l^H\mathbf{w}_l = \textnormal{Tr}(\mathbf{A}^*)$ regardless of the specific realization of $\mathbf{e}_l$.

When $\textnormal{rank}(\mathbf{A}^*) > 1$, it is likely that at least one of the constraints will be violated. However, a feasible weight vector can be obtained by scaling $\mathbf{w}_l$ to satisfy all the constraints. The randomization procedure is detailed in Algorithm \ref{RandomizationAlgorithm}. Additionally, we repeat the random sampling $L$ times and select the vector that provides the optimal value for the objective function.
\begin{algorithm}
\caption{Randomization method \cite{Sidiropoulos2006}.}
\label{RandomizationAlgorithm}
\begin{algorithmic}[1]
\renewcommand{\algorithmicrequire}{\textbf{Input:}}
\renewcommand{\algorithmicensure}{\textbf{Output:}}
\REQUIRE $\mathbf{A}^*$, $L$
\STATE Calculate eigen-decomposition $\mathbf{A}^* = \mathbf{U} \mathbf{\Sigma} \mathbf{U}^H$;
\FOR {$l = 1$ to $L$}
\STATE Generate $[\mathbf{e}_l]_i = e^{j \theta _{l,i}}$ where $\theta _{l,i}\sim \mathcal{U}[0,2\pi)$;
\STATE Construct a feasible point $\tilde{\mathbf{w}}_l = \mathbf{U} \mathbf{\Sigma}^{1/2} \mathbf{e}_l$;
\ENDFOR
\STATE Determine $l^* = \arg \min_{l= 1,...,L} \left \| \mathbf{H}_E \tilde{\mathbf{w}}_l\right \|_2^2$;
\ENSURE  $\hat{\mathbf{w}} = \tilde{\mathbf{w}}_{l^*}$ 
\end{algorithmic} 
\end{algorithm}

\section{SEP-based M-ary Beamforming}	
\label{M_AryBeamformingProblem}
\subsection{Problem Formulation}
This section extends our analysis to M-ary detection schemes with $M$ distinct transmit signals, addressing high-bit rate scenarios. Unlike the binary antipodal detection in Section \ref{BinaryAntipodalBeamformingProblem}, M-ary detection enhances spectral efficiency for more complex communication systems. Consequently, the union bound on the SEP for both Bob and Eve, derived from \eqref{UnionBoundSymbolErrorProbabilityFormualtion}, provides a comprehensive framework for evaluating beamforming performance in high-bit-rate environments as follows
\begin{align}
\bar{P_e}^B &= \frac{1}{M} \sum_{i=1}^{M} \sum_{j=1, j \neq i}^{M} Q\left ( \frac{\parallel  \mathbf{H}_B \mathbf{W}(\mathbf{s}_i - \mathbf{s}_j)\parallel_2 }{2\sqrt{N_B/2}} \right ), \\
\bar{P_e}^E &= \frac{1}{M} \sum_{i=1}^{M} \sum_{j=1, j \neq i}^{M} Q\left ( \frac{\parallel  \mathbf{H}_E \mathbf{W}(\mathbf{s}_i - \mathbf{s}_j)\parallel_2 }{2\sqrt{N_E/2}} \right ),
\end{align}

From (\ref{SymbolErrorProbabilityFormualtion}), we can derive the lower bound of the SEP for M-ary detection as below
\begin{eqnarray}
P_e^{M-ary} &=& \sum_{i = 1}^M P(\hat{\mathbf{s}} \neq \mathbf{s}_i, \mathbf{s} = \mathbf{s}_i),\\
&=& \sum_{i = 1}^M P(\hat{\mathbf{s}} \neq \mathbf{s}_i| \mathbf{s} = \mathbf{s}_i) P(\mathbf{s} = \mathbf{s}_i), \\
&=& \frac{1}{M} \sum_{i = 1}^M P(\hat{\mathbf{s}} \neq \mathbf{s}_i| \mathbf{s} = \mathbf{s}_i), \\
&\geq& \frac{1}{M} M \min_i P(\hat{\mathbf{s}} \neq \mathbf{s}_i| \mathbf{s} = \mathbf{s}_i), \\
&\geq& \min_{i \in \{1,...,M\}} P(\hat{\mathbf{s}} \neq \mathbf{s}_i| \mathbf{s} = \mathbf{s}_i), \\
&\geq& \min_{i,j \in \{1,...,M\}, i \neq j} P(\hat{\mathbf{s}} = \mathbf{s}_j| \mathbf{s} = \mathbf{s}_i),
\label{SEP_Eve_LB}
\end{eqnarray}
where $P(\hat{\mathbf{s}} = \mathbf{s}_j| \mathbf{s} = \mathbf{s}_i) = Q\left ( \frac{\left \| \mathbf{H} \mathbf{W} (\mathbf{s}_i - \mathbf{s}_j) \right \|_2}{2\sqrt{N_0/2}} \right )$.

Let $P_{LB}^E$ be the lower bound of SEP for Eve in M-ary signal detection. From (\ref{SEP_Eve_LB}), we can denote as 
\begin{eqnarray}
P_{LB}^E &=& \min_{i, j \in \{1, ..., M\}} P(\hat{\mathbf{s}} = \mathbf{s}_j| \mathbf{s} = \mathbf{s}_i), \\
&=& \min_{i, j \in \{1, ..., M\}, i \neq j} Q\left ( \frac{\left \| \mathbf{H}_E \mathbf{W} (\mathbf{s}_i - \mathbf{s}_j) \right \|_2}{2\sqrt{N_E/2}} \right ),
\end{eqnarray}

In this problem, our objective is to minimize Bob's SEP bound while ensuring that $P_{LB}^E$ exceeds a predetermined threshold, all within the given power constraints. The optimization problem for achieving this secrecy can be formulated as follows:
\begin{mini!}|s|[2] 
{\mathbf{W}} 
{\frac{1}{M} \sum_{i=1}^{M} \sum_{j=1, j \neq i}^{M} Q\left ( \frac{\parallel  \mathbf{H}_B \mathbf{W}(\mathbf{s}_i - \mathbf{s}_j)\parallel_2 }{2\sqrt{N_B/2}} \right ) } 
{\label{M-aryProb}} 
{} 
\addConstraint{\min_{i, j \in \{1, ..., M\}, i \neq j} Q\left ( \frac{\left \| \mathbf{H}_E \mathbf{W} (\mathbf{s}_i- \mathbf{s}_j) \right \|_2}{2\sqrt{N_E/2}} \right ) \geq H,} 
\addConstraint{\textnormal{Tr} (\mathbf{W}\mathbf{W}^H)}{\leq P, \label{}} 
\end{mini!}	
where $\mathbf{H}_B \in \mathbb{C}^{K_B\times N}$, $\mathbf{H}_E \in \mathbb{C}^{K_E\times N}$, $\mathbf{W} \in \mathbb{C}^{N\times L}$, $ \left \{ \mathbf{s}_1,..., \mathbf{s}_M \right \}  \in \mathbb{C}^{L}$, $H > 0$, and $P > 0$.

To facilitate the analysis, it turns out convenient to transform the complex-valued problem (\ref{M-aryProb}) into an equivalent real-valued problem using the following definitions:
\begin{align}
\tilde{\mathbf{W}} &= \begin{bmatrix}
\Re\left \{ \mathbf{W} \right \} & - \Im \left \{ \mathbf{W} \right \}\\ 
\Im \left \{ \mathbf{W} \right \} & \Re\left \{ \mathbf{W} \right \}
\end{bmatrix}, \\
\tilde{\mathbf{H}}_{B} &= \begin{bmatrix}
\Re\left \{ \mathbf{H}_B \right \} & - \Im \left \{ \mathbf{H}_B \right \}\\ 
\Im \left \{ \mathbf{H}_B \right \} & \Re\left \{ \mathbf{H}_B \right \}
\end{bmatrix},\\
\tilde{\mathbf{H}}_{E} &= \begin{bmatrix}
\Re\left \{ \mathbf{H}_E \right \} & - \Im \left \{ \mathbf{H}_E \right \}\\ 
\Im \left \{ \mathbf{H}_E \right \} & \Re\left \{ \mathbf{H}_E \right \}
\end{bmatrix},\\
\tilde{\mathbf{s}}_i &= \begin{bmatrix}
\Re \left \{ \mathbf{s}_i \right \}\\ 
\Im \left \{ \mathbf{s}_i\right \}
\end{bmatrix},
\end{align}

These definitions allow us to reformulate (\ref{M-aryProb}) as the following real-valued problem:
\begin{mini!}|s|[2] 
{\tilde{\mathbf{W}}} 
{\frac{1}{M} \sum_{i=1}^{M} \sum_{j=1, j \neq i}^{M} Q\left ( \frac{\parallel  \tilde{\mathbf{H}}_B \tilde{\mathbf{W}}(\tilde{\mathbf{s}}_i - \tilde{\mathbf{s}}_j)\parallel_2 }{2\sqrt{N_B/2}} \right ) } 
{\label{UnionBoundProblem}} 
{} 
\addConstraint{\min_{i, j \in \{1, ..., M\}, i \neq j} Q\left ( \frac{\left \| \tilde{\mathbf{H}}_E \tilde{\mathbf{W}} (\tilde{\mathbf{s}}_i - \tilde{\mathbf{s}}_j) \right \|_2}{2\sqrt{N_E/2}} \right ) \geq H,} 
\addConstraint{\textnormal{Tr} (\tilde{\mathbf{W}}\tilde{\mathbf{W}}^T)}{\leq P, \label{ProjectionConstraint}} 
\end{mini!}
where $\tilde{\mathbf{H}}_B \in \mathbb{R}^{2K_B\times 2N}$, $\tilde{\mathbf{H}}_E  \in \mathbb{R}^{2K_E\times 2N}$, $\tilde{\mathbf{W}} \in \mathbb{R}^{2N\times 2L}$, $\left \{ \tilde{\mathbf{s}}_1,..., \tilde{\mathbf{s}}_M \right \}  \in \mathbb{R}^{2L}$.

For given $\tilde{\mathbf{W}}$, we define $i^*$ and $j^*$ as follows:
\begin{equation}
i^*, j^* = \arg \min_{i, j \in \{1, ..., M\}, i \neq j} Q\left ( \frac{\left \| \tilde{\mathbf{H}}_E \tilde{\mathbf{W}} (\tilde{\mathbf{s}}_i - \tilde{\mathbf{s}}_j) \right \|_2}{2\sqrt{N_E/2}} \right ),
\end{equation} 

Due to the non-convexity of Problem \eqref{UnionBoundProblem}, solving it with low-complexity methods is challenging, making the optimal solution difficult to find. To address this, we propose an approximate solution for the reformulated Problem \eqref{ApproximaProblem} with $\gamma > 0$ as a tuning parameter. Hence, $\gamma$ can be tuned until Eve's desired SNR is obtained. This approach simplifies the optimization process while balancing the minimization of Bob’s SEP and ensuring the minimization of Eve’s SNR. The parameter $\gamma$ serves as a trade-off, allowing the solution to meet practical constraints and performance requirements.
\begin{figure*}
\begin{mini!}|s|[2] 
{\tilde{\mathbf{W}}} 
{\frac{1}{M} \sum_{i=1}^{M} \sum_{j=1, j \neq i}^{M} Q\left ( \frac{\parallel  \tilde{\mathbf{H}}_{B} \tilde{\mathbf{W}}(\tilde{\mathbf{s}}_i - \tilde{\mathbf{s}}_j)\parallel_2 }{2\sqrt{N_B/2}} \right ) - \gamma \min_{i, j \in \{1, ..., M\}, i \neq j} Q\left ( \frac{\left \| \tilde{\mathbf{H}}_E \tilde{\mathbf{W}} (\tilde{\mathbf{s}}_i - \tilde{\mathbf{s}}_j) \right \|_2}{2\sqrt{N_E/2}} \right ),}  
{\label{ApproximaProblem}}
{} 
{} 
\addConstraint{\textnormal{Tr} (\tilde{\mathbf{W}} \tilde{\mathbf{W}}^T)}{\leq P, \label{}} 
\end{mini!}	
\noindent\rule{\textwidth}{0.6pt}
\end{figure*}

\begin{figure*}
\begin{eqnarray}
f(\tilde{\mathbf{W}}, \gamma) &=& \frac{1}{M} \sum_{i=1}^{M} \sum_{j=1, j \neq i}^{M} Q\left ( \frac{\parallel  \tilde{\mathbf{H}}_{B} \tilde{\mathbf{W}}(\tilde{\mathbf{s}}_{i} - \tilde{\mathbf{s}}_j)\parallel_2 }{2\sqrt{N_B/2}} \right ) - \gamma Q\left ( \frac{\parallel  \tilde{\mathbf{H}}_{E} \tilde{\mathbf{W}}(\tilde{\mathbf{s}}_{i^*} - \tilde{\mathbf{s}}_{j^*})\parallel_2 }{2\sqrt{N_E/2}} \right ), \\
&=& \frac{1}{M} \sum_{i=1}^{M} \sum_{j=1, j \neq i}^{M} Q\left ( \sqrt{ \frac{\parallel  \tilde{\mathbf{H}}_{B} \tilde{\mathbf{W}}(\tilde{\mathbf{s}}_i - \tilde{\mathbf{s}}_j)\parallel_2 ^2 }{2N_B} } \right )
- \gamma Q\left ( \sqrt{ \frac{\parallel  \tilde{\mathbf{H}}_{E} \tilde{\mathbf{W}}(\tilde{\mathbf{s}}_{i^*} - \tilde{\mathbf{s}}_{j^*})\parallel_2 ^2 }{2N_E} } \right ),\\ 
&=& \frac{1}{M} \sum_{i=1}^{M} \sum_{j=1, j \neq i}^{M} Q\left ( \sqrt{ \frac{ (\tilde{\mathbf{s}}_i - \tilde{\mathbf{s}}_j)^T \tilde{\mathbf{W}}^T \tilde{\mathbf{H}}_{B}^T \tilde{\mathbf{H}}_{B}\tilde{\mathbf{W}} (\tilde{\mathbf{s}}_i - \tilde{\mathbf{s}}_j)  }{2N_B} } \right )
- \gamma Q\left ( \sqrt{ \frac{ (\tilde{\mathbf{s}}_{i^*} - \tilde{\mathbf{s}}_{j^*})^T \tilde{\mathbf{W}}^T \tilde{\mathbf{H}}_{E}^T \tilde{\mathbf{H}}_{E}\tilde{\mathbf{W}} (\tilde{\mathbf{s}}_{i^*} - \tilde{\mathbf{s}}_{j^*})  }{2N_E} } \right ), \nonumber\\
\label{ObjectiveEq}
\end{eqnarray}  
\noindent\rule{\textwidth}{0.6pt}
\end{figure*}

\begin{figure*}
\begin{equation}
\label{gradienObjectiveEq}
\begin{split}
\nabla_{\tilde{\mathbf{W}}} f (\tilde{\mathbf{W}}, \gamma) & =  -\frac{1}{M} \sum_{i=1}^{M} \sum_{j=1, j \neq i}^{M} \! \frac{1}{\sqrt{2\pi}}e^{-\frac{ (\tilde{\mathbf{s}}_i - \tilde{\mathbf{s}}_j)^T \tilde{\mathbf{W}}^T \tilde{\mathbf{H}}_{B}^T \tilde{\mathbf{H}}_{B} \tilde{\mathbf{W}} (\tilde{\mathbf{s}}_i - \tilde{\mathbf{s}}_j)  }{4N_B}} \! \frac{1}{2\sqrt{\frac{ (\tilde{\mathbf{s}}_i - \tilde{\mathbf{s}}_j)^T \tilde{\mathbf{W}}^T \tilde{\mathbf{H}}_{B}^T \tilde{\mathbf{H}}_{B} \tilde{\mathbf{W}} (\tilde{\mathbf{s}}_i - \tilde{\mathbf{s}}_j)  }{2N_B}}} \nabla_{\tilde{\mathbf{W}}} \!\! \left ( \frac{ (\tilde{\mathbf{s}}_i \! - \!\tilde{\mathbf{s}}_j)^T \tilde{\mathbf{W}}^T \tilde{\mathbf{H}}_{B}^T \tilde{\mathbf{H}}_{B} \tilde{\mathbf{W}} (\tilde{\mathbf{s}}_i \! - \! \tilde{\mathbf{s}}_j)  }{2N_B} \right )\\
&+ \gamma \frac{1}{\sqrt{2\pi}}e^{-\frac{ (\tilde{\mathbf{s}}_{i^*} - \tilde{\mathbf{s}}_{j^*})^T \tilde{\mathbf{W}}^T \tilde{\mathbf{H}}_{E}^T \tilde{\mathbf{H}}_{E} \tilde{\mathbf{W}} (\tilde{\mathbf{s}}_{i^*} - \tilde{\mathbf{s}}_{j^*})  }{4N_E}} \! \frac{1}{2\sqrt{\frac{ (\tilde{\mathbf{s}}_{i^*} - \tilde{\mathbf{s}}_{j^*})^T \tilde{\mathbf{W}}^T \tilde{\mathbf{H}}_{E}^T \tilde{\mathbf{H}}_{E} \tilde{\mathbf{W}} (\tilde{\mathbf{s}}_{i^*} - \tilde{\mathbf{s}}_{j^*})  }{2N_E}}} \nabla_{\tilde{\mathbf{W}}} \!\! \left ( \frac{ (\tilde{\mathbf{s}}_{i^*} \! - \!\tilde{\mathbf{s}}_{j^*})^T \tilde{\mathbf{W}}^T \tilde{\mathbf{H}}_{E}^T \tilde{\mathbf{H}}_{E} \tilde{\mathbf{W}} (\tilde{\mathbf{s}}_{i^*} \! - \! \tilde{\mathbf{s}}_{j^*})  }{2N_E} \right ), \\
&= -\frac{1}{M} \sum_{i=1}^{M} \sum_{j=1, j \neq i}^{M} \frac{1}{\sqrt{2\pi}}e^{-\frac{ (\tilde{\mathbf{s}}_i - \tilde{\mathbf{s}}_j)^T \tilde{\mathbf{W}}^T \tilde{\mathbf{H}}_{B}^T \tilde{\mathbf{H}}_{B} \tilde{\mathbf{W}} (\tilde{\mathbf{s}}_i - \tilde{\mathbf{s}}_j)  }{4N_B}}\frac{1}{2\sqrt{\frac{ (\tilde{\mathbf{s}}_i - \tilde{\mathbf{s}}_j)^T \tilde{\mathbf{W}}^T \tilde{\mathbf{H}}_{B}^T \tilde{\mathbf{H}}_{B} \tilde{\mathbf{W}} (\tilde{\mathbf{s}}_i - \tilde{\mathbf{s}}_j)  }{2N_B}}} \frac{\tilde{\mathbf{H}}_{B}^T \tilde{\mathbf{H}}_{B} \tilde{\mathbf{W}} (\tilde{\mathbf{s}}_i - \tilde{\mathbf{s}}_j) (\tilde{\mathbf{s}}_i - \tilde{\mathbf{s}}_j)^T}{N_B} \\
&+ \gamma \frac{1}{\sqrt{2\pi}}e^{-\frac{ (\tilde{\mathbf{s}}_{i^*} - \tilde{\mathbf{s}}_{j^*})^T \tilde{\mathbf{W}}^T \tilde{\mathbf{H}}_{E}^T \tilde{\mathbf{H}}_{E} \tilde{\mathbf{W}} (\tilde{\mathbf{s}}_{i^*} - \tilde{\mathbf{s}}_{j^*})  }{4N_E}}\frac{1}{2\sqrt{\frac{ (\tilde{\mathbf{s}}_{i^*} - \tilde{\mathbf{s}}_{j^*})^T \tilde{\mathbf{W}}^T \tilde{\mathbf{H}}_{E}^T \tilde{\mathbf{H}}_{E} \tilde{\mathbf{W}} (\tilde{\mathbf{s}}_{i^*} - \tilde{\mathbf{s}}_{j^*})  }{2N_E}}} \frac{\tilde{\mathbf{H}}_{E}^T \tilde{\mathbf{H}}_{E} \tilde{\mathbf{W}} (\tilde{\mathbf{s}}_{i^*} - \tilde{\mathbf{s}}_{j^*}) (\tilde{\mathbf{s}}_{i^*} - \tilde{\mathbf{s}}_{j^*})^T}{N_E}, 
\end{split}
\end{equation}
\noindent\rule{\textwidth}{0.6pt}
\end{figure*}

\subsection{SEP--M-ary Beamforming Algorithm}
We propose the projected-gradient-descent (PGD)-based algorithm to solve the Problem (\ref{ApproximaProblem}). PGD algorithm is a first-order optimization technique designed to solve constrained optimization problems. The method involves projecting the gradient of the objective function onto the constraint set and subsequently taking a step in the direction of this projected gradient \cite{Boyd2004}.

Given $\tilde{\mathbf{W}}$ at a specific step, we can determine $i^*, j^* = \arg \min_{i, j \in \{1, ..., M\}, i \neq j} Q\left ( \frac{\left \| \tilde{\mathbf{H}}_E \tilde{\mathbf{W}} (\tilde{\mathbf{s}}_i - \tilde{\mathbf{s}}_j) \right \|_2}{2\sqrt{N_E/2}} \right )$. Therefore, the objective function of Problem \eqref{ApproximaProblem} is defined as Equation (\ref{ObjectiveEq}), and its gradient is derived in Equation (\ref{gradienObjectiveEq}). The PGD method is outlined as follows:
\begin{equation}
\label{ProjectEq}
\tilde{\mathbf{W}}^{(k+1)}\!=\!\textnormal{Proj}_{\left \{ \textnormal{Tr} (\tilde{\mathbf{W}}^{(k)}(\tilde{\mathbf{W}}^{(k)})^T) \leq P \right \}} \!\!\left (  \tilde{\mathbf{W}}^{(k)} \!-\! \alpha \nabla_{\tilde{\mathbf{W}}} f(\tilde{\mathbf{W}}^{(k)}, \gamma) \right )\!\!,
\end{equation}

In PGD, the goal is to minimize a function while ensuring that the iterates remain within a certain feasible set (i.e., a set that satisfies certain constraints).

{\bf Gradient Step.} Taking a step in the direction of the gradient.
\begin{equation}
\mathbf{G}=\tilde{\mathbf{W}}^{(k)} - \alpha \nabla_{\tilde{\mathbf{W}}} f(\tilde{\mathbf{W}}^{(k)}, \gamma),
\end{equation}
where $\alpha$ denotes the step size taken in the direction of the negative gradient.

{\bf Projection Step.} Project the gradient descent update onto the feasible set (i.e., $\textnormal{Tr} (\tilde{\mathbf{W}}^{(k)}(\tilde{\mathbf{W}}^{(k)})^T) \leq P$). This step ensures that the iterates remain feasible.
\begin{eqnarray}
\tilde{\mathbf{W}}^{(k+1)} &=&\textnormal{Proj}_{\left \{ \textnormal{Tr} (\tilde{\mathbf{W}}^{(k)}(\tilde{\mathbf{W}}^{(k)})^T) \leq P \right \}}\left (  \mathbf{G} \right ), \\
& =& \left\{\begin{matrix}
\arg \min_{\tilde{\mathbf{W}}_{\textnormal{Proj}}} \left \| \mathbf{G} - \tilde{\mathbf{W}}_{\textnormal{Proj}} \right \|_F, \\ \\
\textnormal{s.t.} \hspace{5 mm} 
\textnormal{Tr} (\tilde{\mathbf{W}}_{\textnormal{Proj}}\tilde{\mathbf{W}}_{\textnormal{Proj}}^T ) \leq P,
\end{matrix}\right. \\
& =& \left\{\begin{matrix}
\arg \min_{\tilde{\mathbf{W}}_{\textnormal{Proj}}} \left \| \mathbf{G} - \tilde{\mathbf{W}}_{\textnormal{Proj}} \right \|_F, \\ \\
\textnormal{s.t.} \hspace{5 mm} 
\left \| \tilde{\mathbf{W}}_{\textnormal{Proj}} \right \|_F \leq  \sqrt{P},
\end{matrix}\right.
\label{projectionGradientDescent}
\end{eqnarray}
	
Let $\mathbf{g} = \textnormal{vec}(\mathbf{G})$ and $\mathbf{w}_{\textnormal{Proj}} = \textnormal{vec}(\tilde{\mathbf{W}}_{\textnormal{Proj}})$. We can then reformulate (\ref{projectionGradientDescent}) as the following problem:
\begin{equation}
\label{CloseFormProjection}
\begin{split}
\tilde{\mathbf{W}}^{(k+1)} = \textnormal{vec}^{-1}(\mathbf{g}^*),
\end{split}
\end{equation}
where
\begin{eqnarray}
\mathbf{g}^* & = & \left\{\begin{matrix}
\arg \min_{\mathbf{w}_{\textnormal{Proj}}} \left \| \mathbf{g} - \mathbf{w}_{\textnormal{Proj}} \right \|_2 \\ \\
\textnormal{s.t.} \hspace{5 mm} 
\left \| \mathbf{w}_{\textnormal{Proj}} \right \|_2 \leq  \sqrt{P},
\end{matrix}\right. \\
&= & \left\{\begin{matrix}
\arg \min_{\mathbf{w}_{\textnormal{Proj}}} \left \| \frac{\mathbf{g}}{\sqrt{P}} - \frac{\mathbf{w}_{\textnormal{Proj}}}{\sqrt{P}} \right \|_2 \\ \\
\textnormal{s.t.} \hspace{5 mm} 
\left \| \frac{\mathbf{w}_{\textnormal{Proj}}}{\sqrt{P}} \right \|_2 \leq  1,
\end{matrix}\right. \\
&= & \begin{cases}
\mathbf{g}\hspace{13.4 mm}  \textnormal{if} \hspace{3 mm}  \left \| \frac{\mathbf{g}}{\sqrt{P}} \right \|_2 \leq 1, \\\\
\sqrt{P}\frac{\mathbf{g}}{\left \| \mathbf{g} \right \|_2} \hspace{3 mm}  \textnormal{if} \hspace{3 mm}  \left \| \frac{\mathbf{g}}{\sqrt{P}} \right \|_2 > 1,
\end{cases}
\end{eqnarray}
	
The method stops if 
\begin{equation}
\label{StopEq}
f(\tilde{\mathbf{W}}^{(k+1)}, \gamma) - f(\tilde{\mathbf{W}}^{(k)}, \gamma) \leq \epsilon, 
\end{equation}
where $\epsilon$ is a very small real number, e.g., $10^{-5}$.
\begin{algorithm}
\caption{SEP--M-ary Beamforming.}
\label{AlgorithmPGD}
\begin{algorithmic}[1]
\renewcommand{\algorithmicrequire}{\textbf{Input:}}
\renewcommand{\algorithmicensure}{\textbf{Output:}}
\REQUIRE $\tilde{\mathbf{W}}^{(0)}$, $P$, $\alpha$, $\gamma$
\ENSURE  $\tilde{\mathbf{W}}^*$
\STATE \textbf{Initialization:} Set $k= 0$
\REPEAT
\STATE $(i^*, j^*) = \arg \min_{i, j \in \{1, ..., M\}, i \neq j} Q\left ( \frac{\left \| \tilde{\mathbf{H}}_E \tilde{\mathbf{W}}^{(k)} (\tilde{\mathbf{s}}_i - \tilde{\mathbf{s}}_j) \right \|_2}{2\sqrt{N_E/2}} \right )$;
\STATE $\mathbf{G}=\tilde{\mathbf{W}}^{(k)} - \alpha \nabla_{\tilde{\mathbf{W}}} f(\tilde{\mathbf{W}}^{(k)}, \gamma);$ \link{\textbf{\small Gradient}} \\  
\STATE $\tilde{\mathbf{W}}^{(k+1)} =\textnormal{Proj}_{\left \{ \textnormal{Tr} (\tilde{\mathbf{W}}^{(k)}(\tilde{\mathbf{W}}^{(k)})^T) \leq P \right \}}\left (  \mathbf{G} \right ),$\\ 
$= \left\{\begin{matrix}
\arg \min_{\tilde{\mathbf{W}}_{\textnormal{Proj}}} \left \| \mathbf{G} - \tilde{\mathbf{W}}_{\textnormal{Proj}} \right \|_F, \\ \\
\textnormal{s.t.} \hspace{5 mm} 
\left \| \tilde{\mathbf{W}}_{\textnormal{Proj}} \right \|_F \leq  \sqrt{P},
\end{matrix}\right. $ \\
$\tilde{\mathbf{W}}^{(k+1)} = \textnormal{vec}^{-1}(\mathbf{g}^*)$;    \link{\textbf{\small Projection}} \\
where
$\mathbf{g}^* = \begin{cases}
\mathbf{g}\hspace{13.4 mm}  \textnormal{if} \hspace{3 mm}  \left \| \frac{\mathbf{g}}{\sqrt{P}} \right \|_2 \leq 1, \\\\
\sqrt{P}\frac{\mathbf{g}}{\left \| \mathbf{g} \right \|_2} \hspace{3 mm}  \textnormal{if} \hspace{3 mm}  \left \| \frac{\mathbf{g}}{\sqrt{P}} \right \|_2 > 1,
\end{cases}$ \\ 
and
$\begin{cases}
\mathbf{g} = \textnormal{vec}(\mathbf{G}), \\
\mathbf{w}_{\textnormal{Proj}} = \textnormal{vec}(\tilde{\mathbf{W}}_{\textnormal{Proj}}),
\end{cases}$
\STATE $k= k+1;$
\UNTIL a stopping criterion is satisfied, i.e., Eq. (\ref{StopEq})
\end{algorithmic} 
\end{algorithm}

The PGD-based method is outlined in Algorithm \ref{AlgorithmPGD}. This algorithm is motivated by its simplicity and memory efficiency, leveraging the benefits of gradient descent. Moreover, the projection onto (\ref{ProjectionConstraint}) has a closed-form solution as shown in (\ref{CloseFormProjection}), allowing for efficient processing of the entire procedure.

\textbf{Time Complexity.} The time complexity of Algorithm \ref{AlgorithmPGD} is evaluated based on its iterative structure. Each iteration involves three primary steps. Step 3 has a complexity of $O(M^2 \times K \times N)$, resulting from the matrix-vector product computations and the exhaustive search over all possible values of $i$ and $j$. Step 4 has a complexity of $O(M^2 \times K \times N \times L)$, which pertains to the gradient calculation of the objective function in Problem \eqref{ApproximaProblem} and goes through all possible values of $i$ and $j$, where $L \leq \min \{N, K\}$. Step 5, which involves the projection procedures, has a complexity of $O(N \times L)$. Consequently, the overall complexity for a single iteration is $O(M^2 \times K \times N \times L)$. Taking into account the total number of iterations required for the algorithm to converge, the overall complexity is $O(\textnormal{number of iterations} \times M^2 \times K \times N \times L)$. This thorough complexity analysis underscores the computational feasibility and efficiency of the proposed algorithm for managing large-scale MIMO systems.

\section{Numerical results}
\label{NumericalResults}
This section presents numerical examples demonstrating the SEP of Gaussian MIMO wiretap channels employing the proposed beamforming algorithms. The signal-to-noise ratio is measured as $SNR = P/N_B$.  

\subsection{Binary Antipodal Beamforming}
\begin{figure}
\begin{center}
\includegraphics[width=0.43\textwidth]{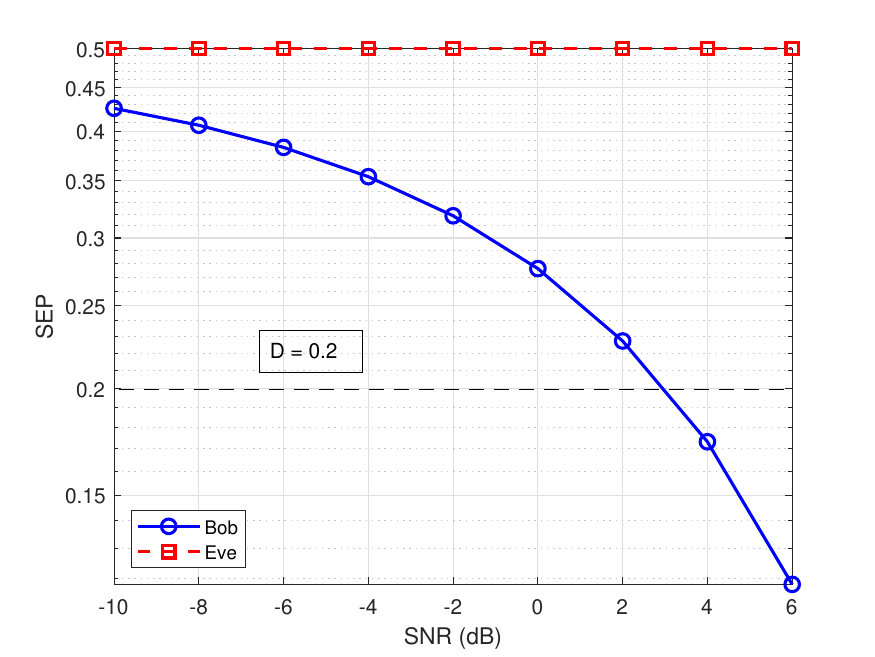}
\caption{SEP comparison of Bob and Eve versus SNR in deterministic real orthogonal direction channels with $N=K_B=K_E = 2$.}	
\label{Numerical1}	        
\end{center}	
\end{figure}

In Figure \ref{Numerical1}, we analyze Bob and Eve's SEP comparison versus SNR, where $N = K_B = K_E = L = 2, a \in \{-1,1\}$, $D = 0.2$, and $N_B = N_E = 0.1$ W. Here, we consider $\mathbf{H}_B = \begin{bmatrix}
0.21 & 0.21 \\ 
0.21 & 0.21
\end{bmatrix}$ and $\mathbf{H}_E = \begin{bmatrix}
0.21 & -0.21 \\ 
-0.21 & 0.21
\end{bmatrix}$ as deterministic orthogonal direction channels, i.e., $\mathbf{H}_B\mathbf{H}_E^T = \mathbf{0}$. Given that the directions of $\mathbf{H}_B$ and $\mathbf{H}_E$ are orthogonal, the optimal beamforming strategy directs the entire information signal to Bob. Consequently, while Eve's error probability $P_e^E$ remains zero, Bob's SEP decreases significantly with increasing SNR.
\begin{figure}
\begin{center}
\includegraphics[width=0.43\textwidth]{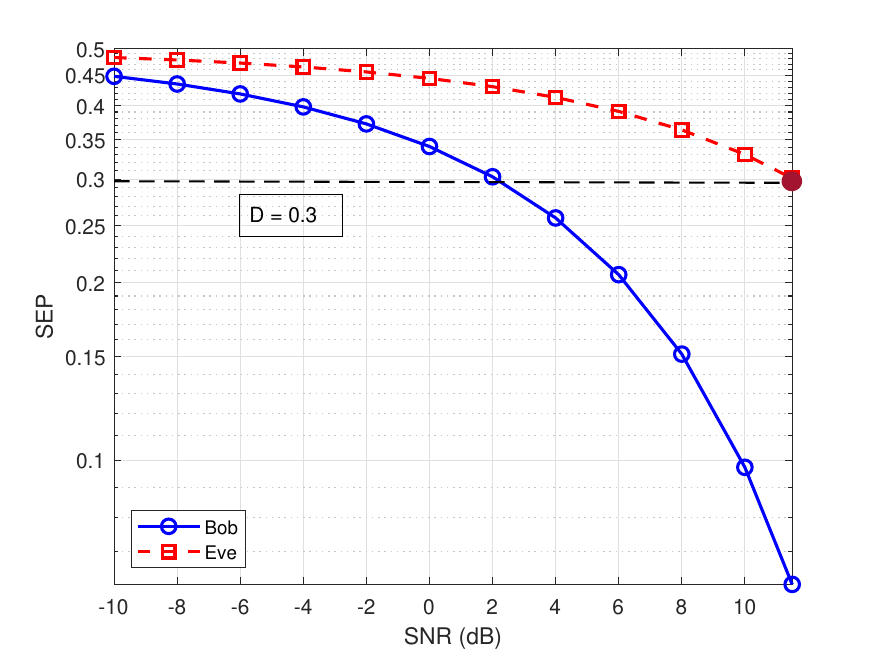}
\caption{SEP comparison of Bob and Eve versus SNR in deterministic real Gaussian channels with $N=K_B=K_E = 2$.}	
\label{Numerical2}	        
\end{center}	
\end{figure}

Figure \ref{Numerical2} illustrates a comparison of SEP for Bob and Eve as a function of SNR, using the same configuration as in Figure \ref{Numerical1}. Here, we set $D = 0.3$, $N_B = N_E = 0.01$ W, and consider $\mathbf{H}_B = \begin{bmatrix}
0.0262 & 0.0049 \\ 
-0.1598 & -0.2414
\end{bmatrix}$ and $\mathbf{H}_E = \begin{bmatrix}
0.0498 & 0.0194 \\ 
-0.0446 & -0.0758
\end{bmatrix}$ as deterministic real Gaussian channels. A notable trend observed for both Bob and Eve is that their performance improves considerably with increasing SNR. Nevertheless, Bob exhibits a lower symbol error probability compared to Eve. This difference arises from the beamforming vector being optimized to direct the information signal towards Bob. It should be noted that constraints (\ref{Case3Constraint1}) and (\ref{Case4Constraint1}) only become active at the brown point. Furthermore, as the SNR increases, none of the scenarios under consideration yield a feasible solution.
\begin{figure}
\begin{center}
\includegraphics[width=0.43\textwidth]{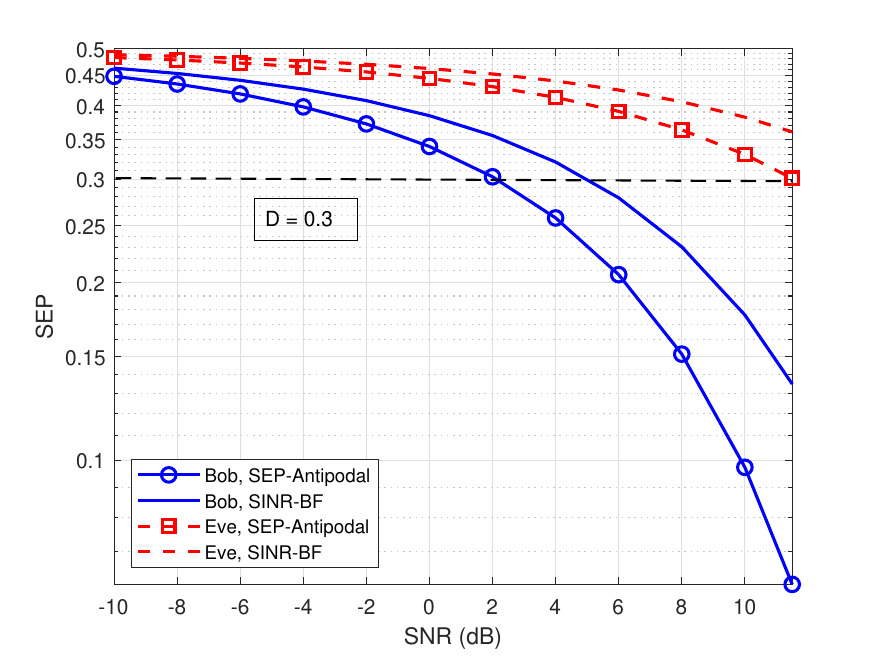}
\caption{SEP versus SNR for Bob and Eve between (i) SEP-Antipodal and (ii) SINR-BF schemes in deterministic real Gaussian channels with $N=K_B=K_E = 2$.}	
\label{Compare1}	        
\end{center}	
\end{figure}

In Figure \ref{Compare1}, we compare the performance of two beamforming schemes: (i) SEP-based binary antipodal beamforming (SEP-Antipodal) scheme, implemented using Algorithm \ref{AlgorithmBeamforming}, and (ii) the signal-to-interference-plus-noise ratio-based beamforming (SINR-BF) scheme, implemented using Algorithm \ref{SINR_Beamforming} in \cite{Mukherjee2011}. The parameter settings are consistent with those used in Figure \ref{Numerical2}. Our scheme significantly outperforms the one proposed in \cite{Mukherjee2011} for both Bob and Eve. For instance, to achieve $P_e^B = 0.25$, SEP-Antipodal requires an SNR of $4$ dB, whereas SINR-BF needs an SNR of $7$ dB. This illustrates the effectiveness of our approach. Furthermore, the performance gap between SEP-Antipodal and SINR-BF widens considerably with increasing SNR.
\begin{algorithm}
\caption{SINR--Beamforming \cite{Mukherjee2011}.}
\label{SINR_Beamforming}
\begin{algorithmic}[1]
\renewcommand{\algorithmicrequire}{\textbf{Input:}}
\renewcommand{\algorithmicensure}{\textbf{Output:}}
\REQUIRE $\mathbf{H}_{B}$, $\mathbf{H}_{E}$, $N$, $K_B$, $K_E$
\ENSURE $\mathbf{w}^*$
\IF{$K_E \leq N$}
    \STATE Compute generalized eigenvector $\mathbf{w}$ corresponding to the largest generalized eigenvalue $\lambda_{\max}$, solving $\mathbf{H}_{B}^H \mathbf{H}_{B}\mathbf{w} = \lambda_{\max} \mathbf{H}_{E}^H \mathbf{H}_{E}\mathbf{w}$;
\ELSIF{$K_E > N$}
    \STATE Compute generalized eigenvector $\mathbf{w}$ corresponding to the smallest generalized eigenvalue $\lambda_{\min}$, solving $\mathbf{H}_{E}^H \mathbf{H}_{E}\mathbf{w} = \lambda_{\min} \mathbf{H}_{B}^H \mathbf{H}_{B}\mathbf{w}$;
\ENDIF
\STATE Scale the vector $\mathbf{w}$ to obtain $\mathbf{w}^*$ such that $\left \| \mathbf{w}^* \right \|_2^2 \leq P$;
\end{algorithmic}
\end{algorithm}

\begin{figure}
\begin{center}
\includegraphics[width=0.43\textwidth]{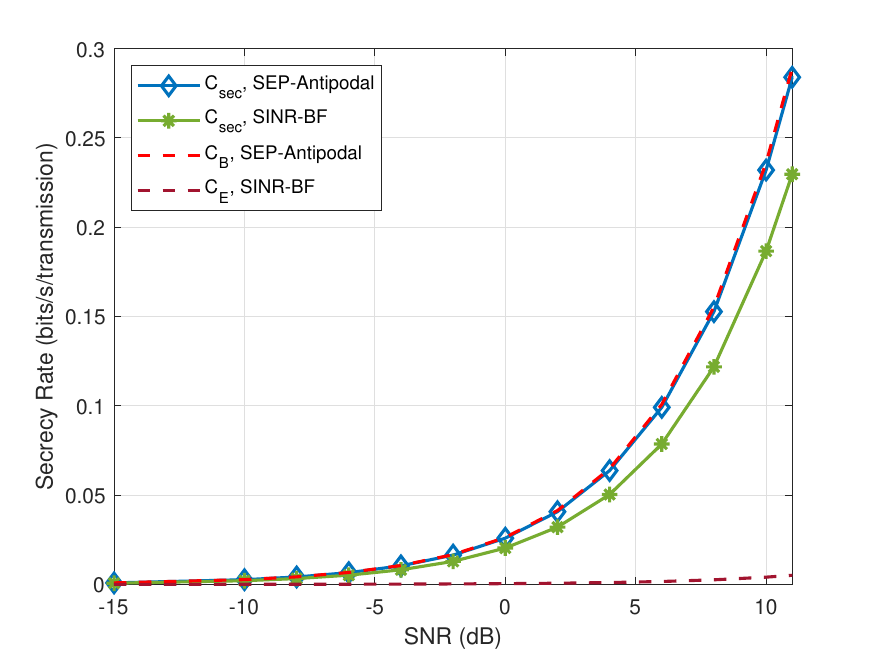}
\caption{Secrecy rate versus SNR between (i) SEP-BF and (ii) SINR-BF schemes in deterministic real Gaussian channels with $N=K_B=K_E = 2$.}	
\label{Compare2}	        
\end{center}	
\end{figure}

\begin{eqnarray}
C_{sec} &=& \max_{\mathbf{Q}_a \succeq 0} C_B - C_E,\\ 
&=& \max_{\mathbf{Q}_a \succeq 0} \log \left | \mathbf{I} + \mathbf{H}_B \mathbf{Q}_a \mathbf{H}_B^H \right | - \log \left | \mathbf{I} + \mathbf{H}_E \mathbf{Q}_a \mathbf{H}_E^H \right |. \nonumber \\
\label{SecrecyCapacity}
\end{eqnarray}
where $\mathbf{Q}_a = \mathbb{E}[\mathbf{w}a(\mathbf{w}a)^H] = a^2 \mathbf{w} \mathbf{w}^H$.

The comparison of secrecy capacity versus SNR between the SEP-Antipodal and SINR-BF schemes is shown in Figure \ref{Compare2}. Secrecy capacity ($C_{sec}$) is determined by Equation (\ref{SecrecyCapacity}). It is observed that the secrecy capacity achieved with SEP-Antipodal surpasses that of SINR-BF, especially in high SNR scenarios. However, both approaches have zero secrecy capacity in the low SNR range. The capacities of Bob ($C_B$) and Eve ($C_E$) corresponding to the SEP-Antipodal case are also illustrated for further detail. It is evident that $C_B$ significantly increases with rising SNR, while $C_E$ remains at zero across the entire SNR range. The configuration is the same as in Figure \ref{Numerical1}, except for $N=K_B=K_E=4$. 
\begin{figure}
\begin{center}
\includegraphics[width=0.43\textwidth]{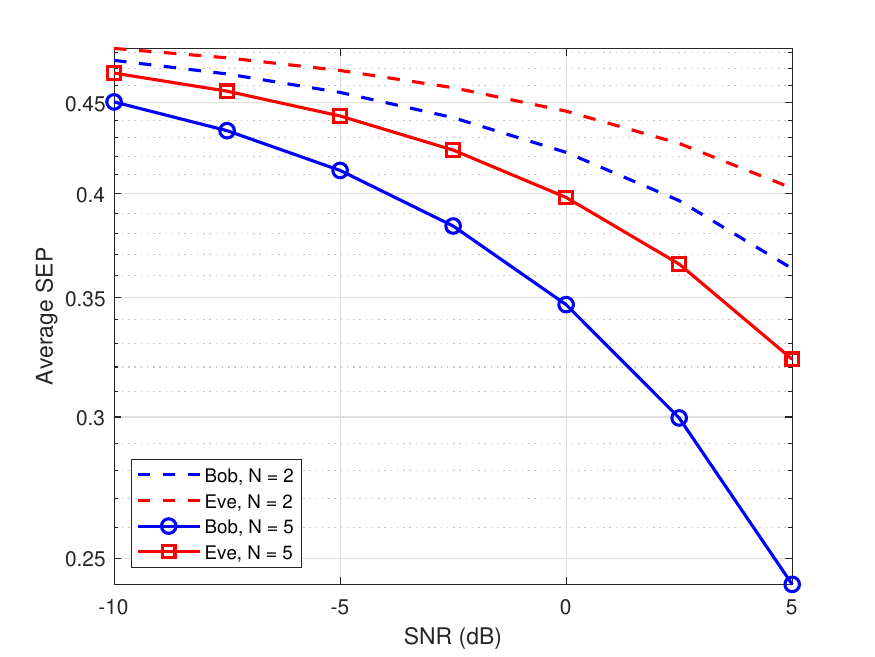}
\caption{Average SEP of Bob and Eve versus SNR with varying $N$ in stochastic real Gaussian channels $K_B=K_E = 2$.}	
\label{Numerical3}	        
\end{center}	
\end{figure} 

\begin{figure}
\begin{center}
\includegraphics[width=0.43\textwidth]{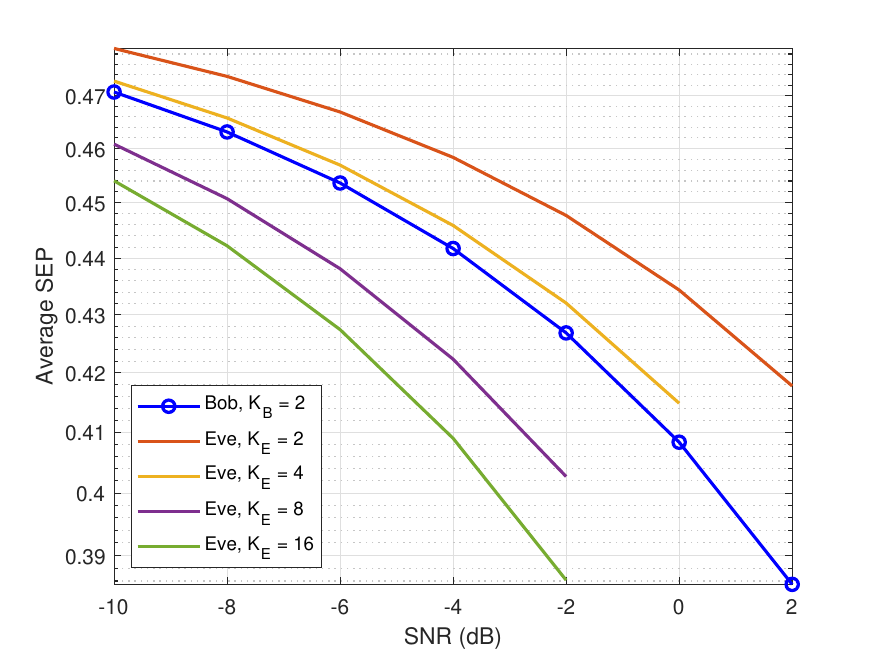}
\caption{Average SEP of Bob and Eve versus SNR with varying $K_E$ in stochastic real Gaussian channels with $N=K_B=2$.}	
\label{Numerical4}	        
\end{center}	
\end{figure}

We investigate the SEP of Bob and Eve as the number of Alice's antennas, $N$, increases in Figure \ref{Numerical4}, using the same system parameters as in Figure \ref{Numerical2}. However, this result is obtained by averaging over 100 realizations of stochastic real Gaussian channels where $h_{ij} \in \mathcal{N}(0, 0.01)$. As shown, the performance of both Bob and Eve improves significantly as $N$ increases from 2 to 5. In addition, the performance gap between Bob and Eve increases as $N$ increases. Hence, increasing the number of transmit antennas helps improve the reliability and security of the communication system. Furthermore, the performance gap between Bob and Eve widens with an increasing $N$. Thus, increasing the number of transmit antennas enhances the reliability and security of the communication system.

The impact of increasing the number of antennas at the eavesdropper is also depicted in Figure \ref{Numerical3}. Eve's average SEP decreases as the number of antennas increases. For $K_E \geq 5$, Eve's performance is notably high, indicating that she can intercept almost all information from Alice. This phenomenon occurs because beamforming is no longer effective at degrading the eavesdropper's reception in this scenario. Thus, information security assurance through physical layer techniques is not guaranteed in such circumstances.
\begin{figure}
\begin{center}
\includegraphics[width=0.43\textwidth]{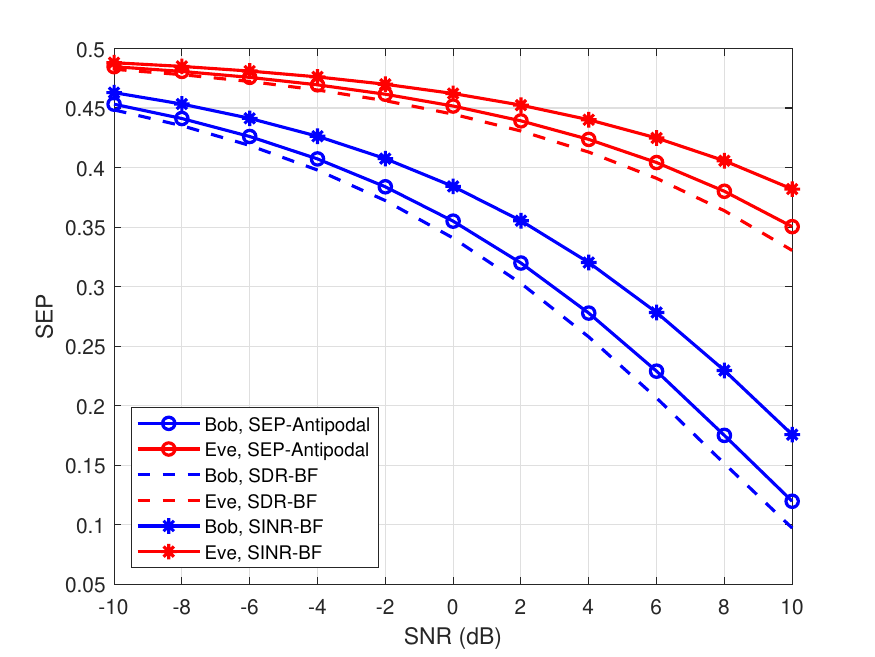}
\caption{SEP versus SNR for Bob and Eve between (i) SEP-Antipodal, (ii) SDR-BF, and (iii) SINR-BF schemes in deterministic real Gaussian channels with $N=K_B=K_E = 2$.}	
\label{Compare3Methods}	        
\end{center}	
\end{figure}

Figure \ref{Compare3Methods} compares the SEP performance of three schemes: (i) SEP-Antipodal, (ii) the SDR-based beamforming scheme (SDR-BF) discussed in Section \ref{OtherProblemFormulation}, and (iii) SINR-BF. Unlike Figure \ref{Compare1}, this one considers the same beamforming vector for each scheme's SNR point. The results indicate that the SEP-Antipodal scheme outperforms SINR-BF, while SDR-BF exhibits slightly better performance than SEP-Antipodal. However, it is important to note that the SDR-BF scheme provides only an approximate solution to the beamforming design problem.   

\subsection{M-ary Beamforming}
In this section, we explore M-ary detection with $M=4$, using the following system parameters: $\mathbf{s}_1 = [1+i \;\; 1-i]^T$, $\mathbf{s}_2 = [-1-i \;\; 1-i]^T$, $\mathbf{s}_3 = [-1+i \;\; 1-i]^T$, $\mathbf{s}_4 = [-1-i \;\; -1+i]^T$, $N = K_B = K_E = L = 2$, and $\gamma = 1$. The channel coefficients are assumed to be deterministic and real. We set the accuracy parameter $\epsilon$ to $10^{-5}$ and allow a maximum of 300 iterations for solving the problem. Each simulation consists of 100 trials, initialized from different starting points.
\begin{figure}
\begin{center}
\includegraphics[width=0.43\textwidth]{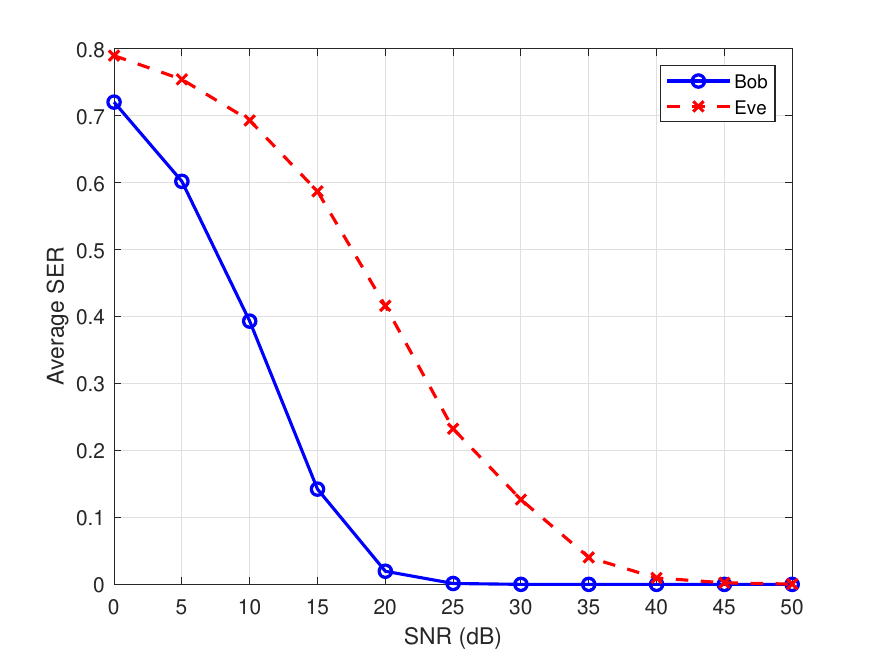}
\caption{Average SER of Bob and Eve versus SNR obtained through Monte Carlo simulation with $N = K_B = K_E = 2$.}	
\label{PGD_ComplexFig1}	        
\end{center}	 
\end{figure}

The Monte Carlo simulation results in Figure \ref{PGD_ComplexFig1} illustrate the SEP of Bob and Eve as a function of SNR. As expected, Bob outperforms Eve across nearly the entire SNR range, thanks to the optimal beamforming matrix. However, at sufficiently high SNR levels, the performance of both Bob and Eve improves significantly. This indicates that physical layer security is not ensured in a very high SNR regime. These results, obtained from actual symbol error rates through Monte Carlo simulation, validate the effectiveness of our upper bound-based approach in practical scenarios.
\begin{figure}
\begin{center}
\includegraphics[width=0.43\textwidth]{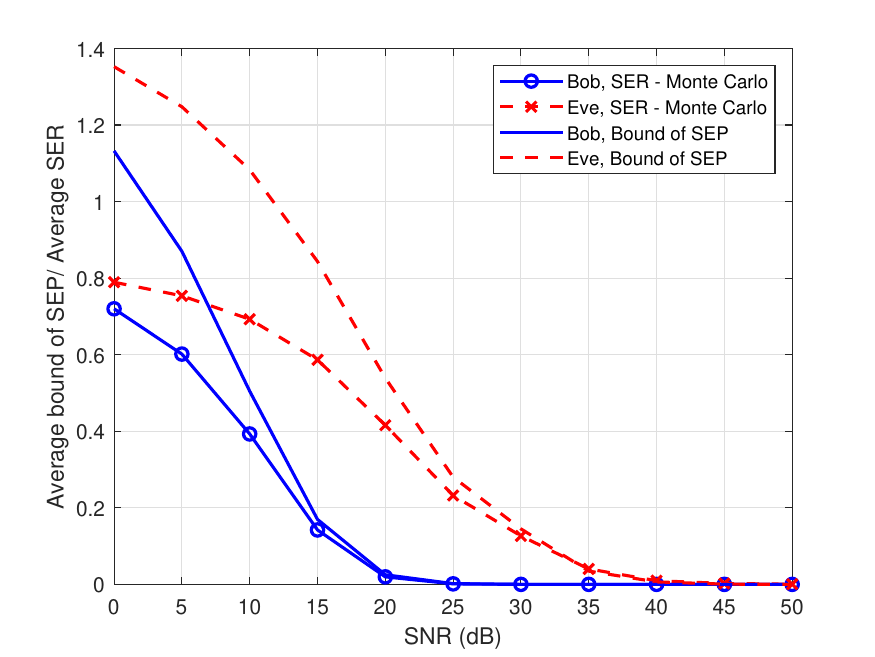}
\caption{Average bound of SEP/ average SER for Bob and Eve versus SNR with $N = K_B = K_E = 2$.}	
\label{PGD_ComplexFig2}	        
\end{center}	 
\end{figure}

Figure \ref{PGD_ComplexFig2} presents the average bounds of SEP, calculated using Algorithm \ref{AlgorithmPGD}, alongside the average SER obtained from Monte Carlo simulation. The results indicate that while the SEP bound is not closely aligned with actual SER values in the low and medium SNR region, it significantly converges in the high SNR range, particularly for $SNR \geq 20$ dB. This suggests that our algorithms, based on the upper bound of SEP, can effectively optimize the beamforming matrix.
\begin{figure}
\begin{center}
\includegraphics[width=0.43\textwidth]{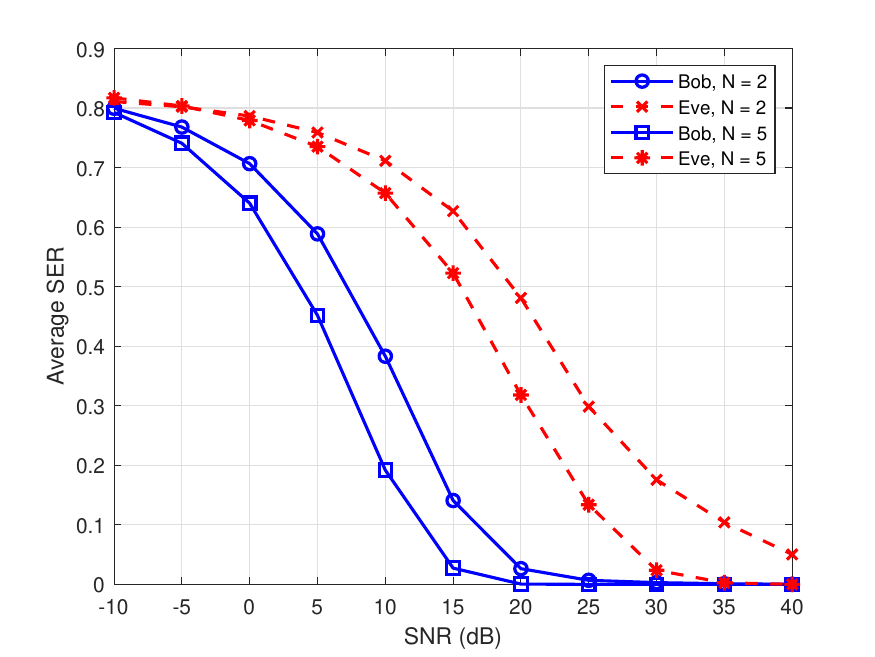}
\caption{Average SER for Bob and Eve versus SNR with different values of $N$ with $K_B = K_E = 2$.}	
\label{PGD_ComplexFig3}	        
\end{center}	 
\end{figure}

\begin{figure}
\begin{center}
\includegraphics[width=0.43\textwidth]{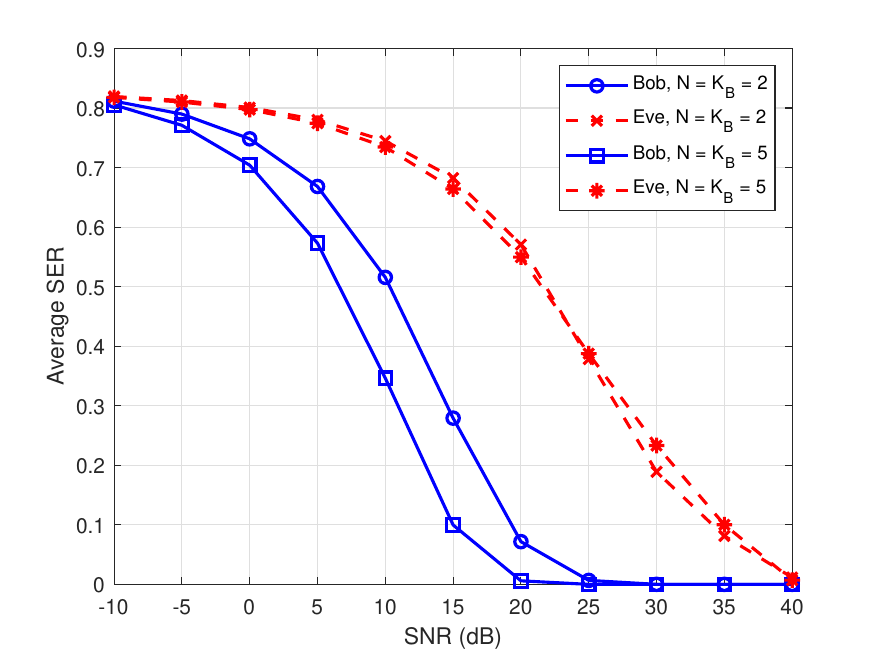}
\caption{Average SER for Bob and Eve versus SNR with different values of $N$ and $K_B$ with $K_E = 2$.}	
\label{PGD_ComplexFig4}	        
\end{center}	 
\end{figure}

In Figure \ref{PGD_ComplexFig3}, we analyze the average SER for both Bob and Eve across different values of $N$, with $K_B = K_E = 2$ set. The results demonstrate a significant improvement in the performance of both Bob and Eve as the number of transmit antennas, $N$, increases. For instance, at an SNR of 10 dB, $P_e^B$ decreases from 0.4 to 0.2 as $N$ increases from 2 to 5. This improvement underscores the importance of increasing the number of transmit antennas to enhance the reliability of the communication system.

The influence of changing $N$ and $K_B$ is explored in Figure \ref{PGD_ComplexFig4}, with $K_E$ fixed at 2. While Bob's SER follows the trend observed in the previous figure, the performance of Eve remains relatively stable between the scenarios with $N=K_B=2$ and $N=K_B=5$. This observation illustrates the benefit of increasing the number of receive antennas to enhance the reliability and security of the MIMO Gaussian wiretap channel.

\section{Conclusion}
\label{Conclusion}
This paper has presented new beamforming schemes for MIMO Gaussian wiretap channels, designed to minimize the symbol error probability for authorized users while constraining the eavesdropper's ability to recover symbols above a predefined threshold. We have proposed an algorithm to determine the optimal beamforming vector for binary antipodal signal detection. Based on Karush-Kuhn-Tucker conditions and the generalized eigen-decomposition method, the proposed algorithm offers the exact solution to the non-convex optimization problem. Furthermore, an approximate and practical algorithm relying on projected gradient descent has been developed to obtain a good beamforming matrix when using M-ary detection schemes. Through extensive numerical simulations, we have demonstrated the efficacy of our approach across various scenarios.




\begin{thebibliography}{99}
\bibitem{Nam2024}
Nam Nguyen, An Vuong, Thuan Nguyen, and Thinh Nguyen,
\enquote{On Minimizing Symbol Error Probability for Antipodal Beamforming in Gaussian MIMO Wiretap Channels,}
\emph{in 2024 IEEE Vehicular Technology Conference,} Washington, D.C., US, Oct. 2024.

\bibitem{Schneier1998}
B. Schneier,
\enquote{Cryptographic design vulnerabilities,}
\emph{IEEE Comput.,} vol. 31, no. 9, pp. 29–33, Sep. 1998.

\bibitem{Mukherjee2014}
A. Mukherjee, S. A. A. Fakoorian, J. Huang, and A. L. Swindlehurst,
\enquote{Principles of physical layer security in multiuser wireless networks: A survey,}
\emph{IEEE Communications Surveys \& Tutorials,} vol. 16, no. 3, pp. 1550–1573, 3rd Quart., 2014.

\bibitem{Liu2017}
Y. Liu, H. -H. Chen and L. Wang,
\enquote{Physical Layer Security for Next Generation Wireless Networks: Theories, Technologies, and Challenges,}
\emph{in IEEE Communications Surveys \& Tutorials,} vol. 19, no. 1, pp. 347-376, Firstquarter 2017.

\bibitem{Wyner1975}
A. D. Wyner,
\enquote{The wire-tap channel,} 
\emph{The Bell System Technical Journal,} vol. 54, no. 8, pp. 1355–1387, 1975.
  
\bibitem{Csiszar1978}
I. Csiszar and J. Korner,
\enquote{Broadcast channels with confidential messages,} 
\emph{in IEEE Transactions on Information Theory,} vol. 24, no. 3, pp. 339-348, May 1978.

\bibitem{Leung1978}
S. Leung-Yan-Cheong and M. Hellman,
\enquote{The Gaussian wire-tap channel,} 
\emph{in IEEE Transactions on Information Theory,} vol. 24, no. 4, pp. 451-456, July 1978.

\bibitem{Goel2005}
S. Goel and R. Negi,
\enquote{Secret communication in presence of colluding eavesdroppers,} 
\emph{in Proc. IEEE Military Commun. Conf.,} Atlantic City, NJ, 2005, pp. 1501–1506.

\bibitem{Khisti2006}
A. Khisti, A. Tchamkerten, and G. W. Wornell,
\enquote{Secure broadcasting with multiuser diversity,} 
\emph{in Proc. IEEE Military Commun. Conf.,} Monticello, IL, Sep. 2006.

\bibitem{Khisti2007}
A. Khisti, G. W. Wornell, A. Wiesel, and Y. Eldar,
\enquote{On the Gaussian MIMO wiretap channel,} 
\emph{presented at the Int. Symp. Inform. Theory,} Nice, France, Jun. 2007.

\bibitem{Li2007}
Z. Li, W. Trappe, and R. Yates,
\enquote{Secret communication via multi-antenna transmission,} 
\emph{presented at the Int. Symp. Inform. Theory,} Baltimore, MD, Mar. 2007.

\bibitem{Gopala2008}
P. Gopala, L. Lai, and H. El Gamal,
\enquote{On the secrecy capacity of fading channels,} 
\emph{IEEE Trans. Inf. Theory,} vol. 54, no. 10, pp. 4687–4698, Oct. 2008.

\bibitem{Liu2009}
T. Liu and S. Shamai,
\enquote{A note on the secrecy capacity of the multiple antenna wiretap channel,} 
\emph{IEEE Transactions on Information Theory,} vol. 55, no. 6, pp. 2547–2553, 2009.

\bibitem{Oggier2011}
F. Oggier and B. Hassibi,
\enquote{The secrecy capacity of the MIMO wiretap channel,} 
\emph{IEEE Transactions on Information Theory,} vol. 57, no. 8, pp. 4961–4972, 2011.

\bibitem{Chakravarty2019}
J. Chakravarty, O. Johnson and R. Piechocki,
\enquote{A Convex Scheme for the Secrecy Capacity of a MIMO Wiretap Channel with a Single Antenna Eavesdropper,} 
\emph{ICC 2019 - 2019 IEEE International Conference on Communications (ICC),} Shanghai, China, 2019, pp. 1-5.

\bibitem{Fakoorian2013}
S. A. A. Fakoorian and A. L. Swindlehurst,
\enquote{Full rank solutions for the MIMO Gaussian wiretap channel with an average power constraint,} 
\emph{IEEE Transactions on Signal Processing,} vol. 61, no. 10, pp. 2620–2631, 2013.

\bibitem{Parada2005}
P. Parada and R. Blahut,
\enquote{Secrecy capacity of SIMO and slow fading channels,} 
\emph{in Proceedings of IEEE International Symposium on Information Theory (ISIT),} pp. 2152–2155, 2005.

\bibitem{Loyka2015}
S. Loyka and C. D. Charalambous,
\enquote{An algorithm for global maximization of secrecy rates in Gaussian MIMO wiretap channels,} 
\emph{IEEE Transactions on Communications,} vol. 63, no. 6, pp. 2288–2299, Jun. 2015.

\bibitem{Li2013}
Q. Li, M. Hong, H.-T. Wai, Y.-F. Liu, W.-K. Ma, and Z.-Q. Luo,
\enquote{Transmit solutions for MIMO wiretap channels using alternating optimization,} 
\emph{IEEE Journal on Selected Areas in Communications,} vol. 31, no. 9, pp. 1714–1727, Sep. 2013.

\bibitem{Nguyen2020}
T. V. Nguyen, Q. -D. Vu, M. Juntti and L. -N. Tran,
\enquote{A Low-Complexity Algorithm for Achieving Secrecy Capacity in MIMO Wiretap Channels,} 
\emph{ICC 2020 - 2020 IEEE International Conference on Communications (ICC),} Dublin, Ireland, 2020, pp. 1-6.  

\bibitem{Dong2020}
L. Dong, S. Loyka and Y. Li,
\enquote{Algorithms for Globally-Optimal Secure Signaling Over Gaussian MIMO Wiretap Channels Under Interference Constraints,} 
\emph{in IEEE Transactions on Signal Processing,} vol. 68, pp. 4513-4528, 2020.

\bibitem{Mukherjee2021}
A. Mukherjee, B. Ottersten and L. -N. Tran,
\enquote{On the Secrecy Capacity of MIMO Wiretap Channels: Convex Reformulation and Efficient Numerical Methods,} 
\emph{in IEEE Transactions on Communications,} vol. 69, no. 10, pp. 6865-6878, Oct. 2021.

\bibitem{Fakoorian2012}
S. A. A. Fakoorian and A. L. Swindlehurst,
\enquote{Optimal power allocation for GSVD-based beamforming in the MIMO Gaussian wiretap channel,} 
\emph{2012 IEEE International Symposium on Information Theory Proceedings,} Cambridge, MA, USA, 2012, pp. 2321-2325.

\bibitem{Khisti2010}
A. Khisti and G. W. Wornell,
\enquote{Secure transmission with multiple antennas–Part II: The MIMOME wiretap channel,} 
\emph{IEEE Transactions on Information Theory,} vol. 56, no. 11, pp. 5515–5532, 2010.

\bibitem{Vaezi2017}
M. Vaezi, W. Shin and H. V. Poor,
\enquote{Optimal Beamforming for Gaussian MIMO Wiretap Channels With Two Transmit Antennas,} 
\emph{IEEE Transactions on Wireless Communications,} vol. 16, no. 10, pp. 6726-6735, Oct. 2017.

\bibitem{Zhang2021}
X. Zhang, Y. Qi, and M. Vaezi,
\enquote{A Rotation-Based Method for Precoding in Gaussian MIMOME Channels,} 
\emph{in IEEE Transactions on Communications,} vol. 69, no. 2, pp. 1189-1200, Feb. 2021.
  
\bibitem{Zhang2019}
X. Zhang and M. Vaezi,
\enquote{Deep Learning Based Precoding for the MIMO Gaussian Wiretap Channel,} 
\emph{2019 IEEE Globecom Workshops (GC Wkshps),} Waikoloa, HI, USA, 2019, pp. 1-6.

\bibitem{Tuan2020}
B. M. Tuan, T. D. Tuyen, N. L. Trung and N. V. Ha,
\enquote{Autoencoder based Friendly Jamming,} 
\emph{2020 IEEE Wireless Communications and Networking Conference (WCNC),} Seoul, Korea (South), 2020, pp. 1-6.

\bibitem{Pauls2022}
J. Pauls and M. Vaezi,
\enquote{Secure Precoding in MIMO-NOMA: A Deep Learning Approach,} 
\emph{in IEEE Wireless Communications Letters,} vol. 11, no. 1, pp. 77-80, Jan. 2022.

\bibitem{Wang2022}
X. Wang, Z. Zheng and Z. Fei,
\enquote{ASAP: Adversarial Learning Based Secure Autoprecoder Design for MIMO Wiretap Channels,} 
\emph{in IEEE Wireless Communications Letters,} vol. 11, no. 9, pp. 1915-1919, Sept. 2022.

\bibitem{Joham2005}
M. Joham, W. Utschick and J. A. Nossek,
\enquote{Linear transmit processing in MIMO communications systems,} 
\emph{in IEEE Transactions on Signal Processing,} vol. 53, no. 8, pp. 2700-2712, Aug. 2005.

\bibitem{Wiesel2006}
A. Wiesel, Y. C. Eldar, and S. Shamai,
\enquote{Linear precoding via conic optimization for fixed MIMO receivers,} 
\emph{in IEEE Transactions on Signal Processing,} vol. 54, no. 1, pp. 161-176, Jan. 2006.

\bibitem{Yu2007}
W. Yu and T. Lan,
\enquote{Transmitter Optimization for the Multi-Antenna Downlink With Per-Antenna Power Constraints,} 
\emph{in IEEE Transactions on Signal Processing,} vol. 55, no. 6, pp. 2646-2660, June 2007.

\bibitem{Mukherjee2011}
A. Mukherjee and A. L. Swindlehurst,
\enquote{Robust Beamforming for Security in MIMO Wiretap Channels With Imperfect CSI,} 
\emph{in IEEE Transactions on Signal Processing,} vol. 59, no. 1, pp. 351-361, Jan. 2011.

\bibitem{Sung2009}
H. Sung, S.. -R. Lee and I. Lee,
\enquote{Generalized Channel Inversion Methods for Multiuser MIMO Systems,} 
\emph{in IEEE Transactions on Communications,} vol. 57, no. 11, pp. 3489-3499, Nov. 2009.

\bibitem{Stankovic2008}
V. Stankovic and M. Haardt,
\enquote{Generalized Design of Multi-User MIMO Precoding Matrices,} 
\emph{in IEEE Transactions on Wireless Communications,} vol. 7, no. 3, pp. 953-961, March 2008.

\bibitem{Hochwald2005}
B. M. Hochwald, C. B. Peel and A. L. Swindlehurst,
\enquote{A vector-perturbation technique for near-capacity multiantenna multiuser communication-part II: perturbation,} 
\emph{in IEEE Transactions on Communications,} vol. 53, no. 3, pp. 537-544, March 2005.

\bibitem{Gallager2008}
Robert G Gallager,
\enquote{Principles of Digital Communication.} 
\emph{Cambridge University Press,} 2008.

\bibitem{Boole1847}
G. Boole,
\enquote{The Mathematical Analysis of Logic.} 
\emph{Philosophical Library,} 1847.

\bibitem{Lopes2023}
E. S. P. Lopes, L. T. N. Landau and A. Mezghani,
\enquote{Minimum Union Bound Symbol Error Probability Precoding for PSK Modulation and Phase Quantization,} 
\emph{2022 IEEE Globecom Workshops (GC Wkshps),}  Rio de Janeiro, Brazil, 2022, pp. 1681-1686.

\bibitem{Christensen2008}
S. S. Christensen, R. Agarwal, E. De Carvalho, and J. M. Cioffi,
\enquote{Weighted sum-rate maximization using weighted MMSE for MIMO-BC beamforming design,} 
\emph{in IEEE Transactions on Wireless Communications,} vol. 7, no. 12, pp. 4792-4799, December 2008.

\bibitem{Yang2013}
N. Yang, P. L. Yeoh, M. Elkashlan, R. Schober and I. B. Collings, 
\enquote{Transmit Antenna Selection for Security Enhancement in MIMO Wiretap Channels,} 
\emph{in IEEE Transactions on Communications,} vol. 61, no. 1, pp. 144-154, January 2013.

\bibitem{Tuy1985}
Tuy, H., T. V. Thieu, and Ng. Q. Thai,
\enquote{A Conical Algorithm for Globally Minimizing a Concave Function over a Closed Convex Set,} 
\emph{Mathematics of Operations Research,} 10, no. 3 (1985): 498–514.

\bibitem{Hoffman1981}
Hoffman, K.L,
\enquote{A method for globally minimizing concave functions over convex sets,} 
\emph{Mathematical Programming,} 20, 22–32 (1981).

\bibitem{Golub1996}
G. Golub and C. F. V. Loan,
\enquote{Matrix Computations,} 
\emph{} 3rd ed. Baltimore, MD: Johns Hopkins Univ. Press, 1996.

\bibitem{Ma2002}
Wing-Kin Ma, T. N. Davidson, Kon Max Wong, Zhi-Quan Luo and Pak-Chung Ching,
\enquote{Quasi-maximum-likelihood multiuser detection using semi-definite relaxation with application to synchronous CDMA,} 
\emph{in IEEE Transactions on Signal Processing,} vol. 50, no. 4, pp. 912-922, April 2002.

\bibitem{Boyd2004}
S. Boyd and L. Vandenberghe,
\enquote{Convex Optimization.} 
\emph{Cambridge, U.K.: Cambridge Univ. Press,} 2004.

\bibitem{Grant2009}
M. Grant, S. Boyd, and Y. Ye. (2009)
\enquote{CVX: MATLAB software for disciplined
convex programming [Online].} Available: http://www.stanford.edu/~boyd/cvx. 

\bibitem{Sidiropoulos2006}
N. D. Sidiropoulos, T. N. Davidson and Zhi-Quan Luo,
\enquote{Transmit beamforming for physical-layer multicasting,} 
\emph{in IEEE Transactions on Signal Processing,} vol. 54, no. 6, pp. 2239-2251, June 2006.

\bibitem{Verdu1998}
S. Verdu,
\enquote{Multiuser Detection.} 
\emph{Cambridge, U.K.: Cambridge University Press}, 1998.

\end{thebibliography}
\end{document}